\documentclass[12pt]{JHEP3}
\usepackage{amsmath}
\usepackage{young}
\usepackage[vcentermath]{youngtab}

\newcommand{\be}{ \begin{equation}}
\newcommand{\ee}{\end{equation}}
\newcommand{\bea}[1]{\begin{eqnarray}\label{#1} }
\newcommand{\eea}{\end{eqnarray}}

\def\ZZZ{{\hskip-3pt\hbox{ Z\kern-1.6mm Z}}}
\def\zzz{{\hskip-3pt\hbox{ z\kern-1mm z}}}

\def\one{{\hbox{ 1\kern-.8mm l}}}
\def\zero{{\hbox{ 0\kern-1.5mm 0}}}

\title{Higher Spins \& Strings}

\author{
Matthias R.\ Gaberdiel$^{a}$ and Rajesh Gopakumar$^{b}$ \\ 
$^a$Institut f\"ur Theoretische Physik, ETH Zurich, \\
$\;$CH-8093 Z\"urich, Switzerland \\
$\;$\email{gaberdiel@itp.phys.ethz.ch}\\ \\ 
$^b$Harish-Chandra Research Institute, \\
$\;$Chhatnag Road, Jhusi,\\
$\;$Allahabad, India 211019\\
$\;$\email{gopakumr@hri.res.in}}

\abstract{It is natural to believe that the free symmetric product orbifold CFT is dual 
to the tensionless limit of string theory on 
${\rm AdS}_3\times {\rm S}^3\times \mathbb{T}^4$.
At this point in moduli space, string
theory is expected to contain a Vasiliev higher spin theory as a subsector.
We confirm this picture explicitly by showing that the large level limit of the 
${\cal N}=4$ cosets of arXiv:1305.4181, that are dual to a higher spin theory on AdS$_3$,
indeed describe a closed subsector of the symmetric product orbifold.
Furthermore, we reorganise the full partition function of the symmetric product orbifold in terms of
representations of the higher spin algebra (or rather its ${\cal W}_{\infty}$ extension). In particular, the 
unbroken stringy symmetries of the tensionless limit are captured by a large chiral algebra
which we can describe explicitly in terms of an infinite sum of ${\cal W}_{\infty}$ representations, thereby
exhibiting a vast extension of the conventional higher spin symmetry.}

\begin{document}

\section{Introduction}

The Vasiliev higher spin theories \cite{Vasiliev:2003ev}
are of great interest to string theorists since they provide a glimpse of the 
mysterious set of enhanced gauge symmetries that string theory has long been presumed to possess. 
In particular, it has often been thought that 
masses in string theory arise from the dynamical breaking of these symmetries, which might be 
restored in special limits. 
Indeed, hints of this enlarged gauge symmetry have been seen in flat space string theory in the high energy (or $\alpha^{\prime}\rightarrow \infty$) regime --- see for instance \cite{Gross:1988ue, Witten:1988zd, Moore:1993qe},
as well as the more recent discussion in \cite{Sagnotti:2011qp}.  
However, it seems plausible that it is in the context of string theory on anti-de sitter (AdS) space times that these 
ideas are fully realised. This is because the AdS background provides an additional length scale --- the radius 
$R$ of AdS ---  and consequently a dimensionless ratio $\frac{R^2}{\alpha^{\prime}}$. We can then imagine taking 
the limit where this ratio goes  uniformly to zero, thereby describing quite plausibly
a genuinely tensionless limit of string theory. 

The AdS/CFT correspondence allows us to make this idea more concrete
\cite{Sundborg:2000wp,Witten,Mikhailov:2002bp,Sezgin:2002rt}. According to the usual
dictionary, the 
dimensionless ratio $\frac{R^2}{\alpha^{\prime}}$ 
is proportional to a coupling or marginal deformation of the dual field theory. Its 
vanishing is therefore equivalent to considering a free gauge theory. A free gauge theory is known to 
have a large set of conserved currents which are, however, not conserved at any non-zero coupling. 
In particular, there are  operators (`twist two' in $d=4$) which are constructed from gauge 
invariant (single trace) bilinears of adjoint valued fields. This set of currents is closed under the OPE, and 
thus forms a consistent subsector \cite{Mikhailov:2002bp}. 
The bulk-boundary dictionary associates to these boundary conserved currents gauge symmetries in AdS. 
In particular, the above universal sector of the free field theory is in nice correspondence with the 
Vasiliev system of interacting gauge fields of spin $s\geq 2$.  

The examples of vector model holography beginning with the work of Klebanov and Polyakov 
\cite{Klebanov:2002ja} (see also \cite{Sezgin:2003pt} for a subsequent generalisation)
have attempted to isolate this sector from the dynamics of the full string theory. For fields transforming 
in the fundamental representation, the only single particle gauge invariant states are bilinears in the fields --- 
the aforementioned universal currents in addition to a few low lying scalars or spin half fields. 
The dynamics of this `perturbative' sector, at leading order in large $N$, is well captured by the classical 
Vasiliev set of equations in the bulk, as confirmed beautifully in the work of Giombi and Yin
\cite{Giombi:2009wh,Giombi:2010vg}. We now also have a compelling scenario for 
embedding these vector-like ${\rm AdS}_4/{\rm CFT}_3$ dualities 
(with SUSY and also Chern-Simons interactions \cite{Aharony:2011jz,Giombi:2011kc}) within 
the ABJ duality for string theory \cite{Chang:2012kt}. 
   
In this paper, we will consider the ${\rm AdS}_3/{\rm CFT}_2$ vector-like dualities of
\cite{Gaberdiel:2010pz} (see \cite{Gaberdiel:2012uj} for a review) in their maximally 
SUSY incarnation \cite{Gaberdiel:2013vva} (see also \cite{Henneaux:2012ny} for 
earlier work), and link them with string theory dualities for ${\rm AdS}_3$, specifically the case of 
${\rm AdS}_3\times {\rm S}^3\times \mathbb{T}^4$. The specific nature of this link seems to be apparently  
different in character from the ${\rm AdS}_4/{\rm CFT}_3$ case, at least to our present understanding. 
The focus in the present work is also somewhat different --- we will try to make a start at addressing the 
issues raised in the opening paragraph in a concrete way, i.e., we aim to explore the enhanced symmetries and 
the nature of their breaking.  We believe the ${\rm AdS}_3/{\rm CFT}_2$ 
case is particularly suited for this endeavour, because, unlike in higher dimensions, the higher spin symmetry group is enlarged here to a ${\cal W}_{\infty}$ algebra \cite{Henneaux:2010xg,Campoleoni:2010zq,Gaberdiel:2011wb}
(generalising the enhancement to Virasoro discovered by Brown and Henneaux \cite{Brown:1986nw}).
Though more complicated than the Virasoro symmetry, 
a number of facts are known about these ${\cal W}_{\infty}$ symmetries and their representations. 
Understanding the stringy symmetries in terms of these 
${\cal W}_{\infty}$ algebras holds the promise of leading to a powerful method that may enable
one to exploit them concretely.
\smallskip

It follows from the general logic of the AdS/CFT correspondence that  string theories on ${\rm AdS}_3$ are 
dual to interacting two dimensional CFTs 
which arise in the IR of  two-dimensional gauge theories describing the near horizon regime of the 
D1-D5 system. Unfortunately, these IR fixed points are not as explicitly understood as in the higher dimensional cases, 
even with maximal supersymmetry. 
For instance, for the case of AdS$_3 \times {\rm S}^{3}\times {\rm S}^{3} \times {\rm S}^1$, there is, 
as of now, no consensus candidate CFT dual (see \cite{Gukov:2004ym}  for a summary of the situation and 
\cite{Tong:2014yna} for a recent proposal; important earlier work is described in 
\cite{Elitzur:1998mm,de Boer:1999rh}). For the better studied case of 
${\rm AdS}_3\times {\rm S}^3\times \mathbb{T}^4$ (see \cite{David:2002wn} for a review and references), the 
dual CFT is believed to be on the moduli space of the free symmetric orbifold  \cite{Dijkgraaf:1998gf}
\be\label{symorb}
{\rm Sym}_{N+1} (\mathbb{T}^4)  \equiv \left. \bigl( \mathbb{T}^4 \bigr)^{N+1}  \right/ S_{N+1} \ ,   
\ee
where $N+1=Q_1Q_5$, the product of D1 and D5 charges. This is the theory which will be the main focus of his 
work, and we will assume for the rest of this paper that it describes indeed the above string compactification.  

Since we do not have a direct gauge theory picture, it is a bit difficult, {\it a priori}, to see what point in the 
moduli space of this CFT might correspond to the tensionless limit. It has often been assumed, though 
never made completely precise to the best of our knowledge, that the free CFT --- of $4(N+1)$ free 
fermions and bosons, subject to the permutation action by $S_{N+1}$ --- is dual to the tensionless 
string theory. We will indeed find very convincing evidence for this picture: in \cite{Gaberdiel:2013vva}
the CFT dual to the ${\cal N}=4$ supersymmetric higher spin theory on AdS$_3$ was identified,
and the sector that captures the `perturbative' Vasiliev states, i.e., the higher spin fields as well 
as the matter multiplets that come with them, was isolated.  Generically, these 
cosets possess the large ${\cal N}=4$ superconformal symmetry, but in the 
limit in which the level of the cosets is taken to infinity, this contracts to the small ${\cal N}=4$ superconformal 
symmetry that controls
the CFT (\ref{symorb}). In this limit, we find that the `perturbative' part of the coset CFT is a very
natural closed subsector of the symmetric orbifold theory, thus mirroring precisely that the perturbative
Vasiliev higher spin theory is a closed subsector of the tensionless string theory.

Given this precise understanding about the relation between  higher spin and string theory, 
we can do much more. In particular, we can organise 
{\it all} the states of the free CFT (\ref{symorb}) into representations of the ${\cal W}_{\infty}[0]$ symmetry 
--- this is the symmetry that controls the infinite level limit of the cosets --- 
with precise, calculable multiplicities. As a special case, we bring some order into the gigantic chiral 
algebra of the symmetric product CFT by identifying the additional ${\cal W}_{\infty}[0]$ representations
that extend ${\cal W}_{\infty}[0]$ --- these correspond to the additional massless higher spin fields
which are present in string theory (but not in the Vasiliev theory, i.e., that are not captured by ${\cal W}_{\infty}$).  
We can also identify them directly in terms of the free bosons and
fermions. This very explicit understanding opens the way towards studying all sorts of 
aspects of this setup, e.g., it should now be possible to explore the higgsing of these fields
under the perturbation that switches on the tension.
\smallskip

In the rest of this introduction, we give a non-technical summary of how these results are obtained. 
We begin in Section~\ref{sec:2} with a brief review of the large ${\cal N}=4$ cosets that are dual
to the supersymmetric higher spin theory on AdS$_3$, concentrating on those aspects that are 
important for the subsequent discussion. These cosets have the same symmetry as string theory on
${\rm  AdS}_3 \times {\rm S}^3 \times {\rm S}^3 \times {\rm S}^1$, where the sizes of the two ${\rm S}^3$'s
correspond to the level $k$ and the rank $N$ of the cosets, respectively. In order to relate them to 
the symmetric orbifold (\ref{symorb}) we then consider in Section~\ref{sec:3} the limit in which
the level $k$ is taken to infinity --- then one of the two ${\rm S}^3$'s decompactifies, and 
we  make
contact with (the zero momentum sector of) string theory on 
${\rm AdS}_3 \times {\rm S}^3 \times \mathbb{T}^4$. In fact, the large ${\cal N}=4$ superconformal
algebra that controls the cosets contracts in this limit to the small ${\cal N}=4$ superconformal
algebra that underlies (\ref{symorb}).  We show in Section~\ref{sec:cosetlim} that in this
limit the `perturbative' part of the CFT cosets can be described in terms of $N+1$ free
bosons and fermions subject to a ${\rm U}(N)$ singlet condition, i.e., as the
untwisted sector of a `continuous orbifold' similar to what was observed for the bosonic case in 
\cite{Gaberdiel:2011aa}. We then show in Section~\ref{sec:3.2} that the 
$S_{N+1}$ permutation action of (\ref{symorb}) is precisely induced from this ${\rm U}(N)$
action via the embedding $S_{N+1}\subset {\rm U}(N)$. It is then immediate that the 
untwisted sector of the continuous orbifold is a natural closed subsector of the untwisted
sector of the symmetric product orbifold (\ref{symorb}) --- thus mirroring precisely that the 
higher spin theory describes a closed subsector of string theory at the tensionless point.

In Section~\ref{sec:comp} we then begin to rewrite the full string spectrum (\ref{symorb}) in terms
of ${\cal W}_{\infty}[0]$ representations; in particular we exhibit in Section~\ref{sec:Wext} how the
full chiral algebra of (\ref{symorb}) can be organised in terms of ${\cal W}_{\infty}[0]$ 
representations and check our answer against explicit predictions of the symmetric orbifold. We also explain
in Section~\ref{4.2.} how the additional symmetry generators may be described in terms of 
the free fermions and bosons.  We finally sketch in Section~\ref{4.3} how a similar analysis may be 
done for the other states of the untwisted sector of (\ref{symorb}). From the viewpoint of the
${\cal W}_\infty[0]$ algebra, the full symmetric orbifold then correponds to a certain non-diagonal
modular invariant.

One important consequence of the extended symmetry that appears in the strin\-gy description
is that the so-called `light states' (that are believed to correspond to classical solutions of
the higher spin theory \cite{Castro:2011iw}) do not define consistent representations of the 
extended symmetry --- the  corresponding space-time interpretation is that these classical solutions 
do not lift to solutions of the full string theory. This is discussed abstractly in Section~\ref{sec:light},
and then more concretely in Section~\ref{sec:7}. (Since these light states can be interpreted
as twisted sectors of the continuous orbifold, we need to explore the dictionary between twisted 
sector states and coset states; this is done in Section~\ref{sec:6}, generalising the recent 
discussion for the ${\cal N}=2$ cosets in \cite{GK}.) As a side-product of the analysis of 
Section~\ref{sec:7} we can also identify in Section~\ref{2cycle} the $2$-cycle twisted states (that contain 
some of the exactly marginal operators of the symmetric orbifold) from the coset point of view; this
explicit description is likely to play an important role in studying the higgsing of the higher spin fields.
Finally, Section~\ref{sec:8} contains our conclusions and an outlook for what sort of questions
could now be addressed. There are a number of appendices (as well as an ancillary file
of the {\tt arXiv} submission) where some of the more technical details are described.

\section{The Large ${\cal N}=4$ Coset}\label{sec:2}

In this section we review the Wolf space coset theories with large ${\cal N}=4$ superconformal symmetry 
\cite{Sevrin:1988ew,Schoutens:1988ig,Spindel:1988sr,Goddard:1988wv,VanProeyen:1989me,Sevrin:1989ce} 
that are dual to a Vasiliev higher spin theory on AdS$_3$ \cite{Gaberdiel:2013vva}.
By taking the large level limit of these cosets we will subsequently make contact 
with the symmetric product theory with target space 
${\rm Sym}_{N+1} (\mathbb{T}^4)$, which has small ${\cal N}=4$ superconformal symmetry, 
and which is known to be dual to string theory on ${\rm AdS}_3\times {\rm S}^3 \times \mathbb{T}^4$. 
 We will restrict ourselves to stating 
those essential results and features, which we shall need in the following sections; for more details, 
we refer the reader to \cite{Gaberdiel:2013vva} and the references contained therein. 
\smallskip

\noindent The Wolf space cosets we will focus on are  given by
\cite{Spindel:1988sr,Goddard:1988wv,VanProeyen:1989me,Sevrin:1989ce}
\be\label{cosetdef}
\frac{\mathfrak{su}(N+2)^{(1)}_{k+N+2}}{\mathfrak{su}(N)^{(1)}_{k+N+2} \oplus \mathfrak{u}(1)^{(1)}_\kappa} 
\oplus \mathfrak{u}(1) 
\ \cong \ \frac{\mathfrak{su}(N+2)_{k}\oplus \mathfrak{so}(4N+4)_1}{\mathfrak{su}(N)_{k+2} \oplus \mathfrak{u}(1)_\kappa}  
\oplus \mathfrak{u}(1)  \ ,
\ee
where the second description is in terms of the bosonic affine algebras. The $\mathfrak{so}$-factor
in the numerator describes $(4N+4)$ real free fermions, and the 
level of the $\mathfrak{u}(1)$ factor in the denominator equals $\kappa= 2N(N+2)(N+k+2)$.
The central charge, given by the usual difference of numerator and denominator central charges, is
\be
c = \frac{6 (N+1)(k+1)}{N+k+2} \ . 
\ee
This coset is known to contain the linear large ${\cal N}=4$ superconformal algebra, whose 
commutation relations are spelt out in Appendix~\ref{app:susy}. For our present purpose, the 
main point to note 
is that it contains in addition to the affine $\mathfrak{u}(1)$ algebra that appears as a direct summand
in (\ref{cosetdef}),  {\it two} $\mathfrak{su}(2)$ affine algebras 
which are at levels $(k+1)$ and $(N+1)$, respectively (see Section~3.1 of \cite{Gaberdiel:2013vva} 
for an explicit construction from the coset viewpoint). The superconformal algebra thus depends,
in addition to the central charge, on a parameter 
$\gamma = \frac{N+1}{N+k+2}$, and is often denoted by $A_{\gamma}$ in the literature. 
The more familiar small ${\cal N}=4$ superconformal algebra is obtained in the limit 
$\gamma \rightarrow 0$ of the above. We shall discuss this limit in more detail in the next section.  

In addition to the large ${\cal N}=4$ superconformal algebra, the coset theory has an extended chiral 
algebra which contains higher spin currents. These can be organised into multiplets $R^{(s)}$ 
of the global superalgebra (with $s=1,2,\ldots$)\footnote{Here and later we will often ignore the 
truncations that occur for finite $N,k$ of the spectrum, since our main interest will require us to take 
both $k$ and $N$ large. However, most of our results remain true at finite $N$, provided
we take $k\rightarrow \infty$.}
\be
\begin{array}{lcc}\label{D2mult2}
& s: & ({\bf 1},{\bf 1})  \\
& s+\tfrac{1}{2}: & ({\bf 2},{\bf 2})  \\
R^{(s)}: \qquad & s+1: & ({\bf 3},{\bf 1}) \oplus ({\bf 1},{\bf 3})  \\
& s+\tfrac{3}{2}: & ({\bf 2},{\bf 2})  \\
& s+2: & \  \  ({\bf 1},{\bf 1})  \ .
\end{array}
\ee
The quantum numbers shown alongside refer to the two $\mathfrak{su}(2)$ algebras. 
This therefore leads to a total of eight currents of a given spin $s=1,\frac{3}{2}, 2, \frac{5}{2},\ldots$ . 
These currents (holomorphic and anti-holomorphic) constitute the vacuum 
sector of the coset theory. They generate a super ${\cal W}_{\infty}[\gamma]$ algebra, which is a generalisation 
of the  somewhat more familiar bosonic ${\cal W}_{\infty}$ algebras. The OPEs (that in effect define
the algebra structure of the ${\cal W}$-algebra) are completely specified by the central charge and the 
parameter $\gamma$ \cite{Beccaria:2014jra},  or equivalently by the levels of the two $\mathfrak{su}(2)$ 
affine algebras, see also \cite{Ahn:2013oya} for a low level analysis.

The states of the CFT can be organised in terms of representations of this ${\cal W}$-symmetry, with
the ${\cal W}$-algebra itself defining the vacuum representation. The most general 
primaries with respect to this  ${\cal W}$-algebra are labelled 
by\footnote{In the following we shall suppress the $\mathfrak{u}(1)$ charge with respect to the numerator 
$\mathfrak{u}(1)$, i.e., all the states we shall consider will be uncharged with respect to this 
$\mathfrak{u}(1)$ algebra. We have furthermore rescaled the $\mathfrak{u}(1)$ charge of the denominator
by a factor of $2$ relative to \cite{Gaberdiel:2013vva} in order to be in line with standard conventions.}
$(\Lambda_+;\Lambda_-,\hat{u})$. This is the usual coset labelling where $\Lambda_+$ is a 
weight of the numerator $\mathfrak{su}(N+2)_k$, while 
$\Lambda_-$ and $\hat{u}\in\mathbb{Z}_\kappa$ are weights of the denominator 
$\mathfrak{su}(N)_{k+2}$  and $\mathfrak{u}(1)_\kappa$ algebra, respectively. These weights 
need to obey a selection rule and determine the primaries up to field identification, as spelled out
in Appendix~\ref{app:susy}. The conformal dimension of the primary $(\Lambda_+;\Lambda_-,\hat{u})$ 
equals
\be\label{N4hdef}
h(\Lambda_+;\Lambda_-,\hat{u}) = \frac{C^{(N+2)}(\Lambda_+)}{N+k+2} - \frac{C^{(N)}(\Lambda_-)}{N+k+2}
- \frac{\hat{u}^2}{4 N (N+2) (N+k+2)} + n \ ,
\ee
where $n$ is a half-integer, describing the `level' at which $(\Lambda_-,\hat{u})$ appears in the representation 
$\Lambda_+$.


\noindent In \cite{Gaberdiel:2013vva} it was proposed that the large $(N,k)$ 't~Hooft limit of these cosets with
\be\label{tH}
\lambda = \frac{N+1}{N+k+2} \quad \hbox{fixed}
\ee
is described by a Vasiliev higher spin theory on AdS$_3$
\cite{Prokushkin:1998bq,Prokushkin:1998vn} with the same supersymmetry. The higher spin 
equations are derived from a theory with local gauge symmetry based on the infinite dimensional gauge algebra
${\rm shs}_2[\lambda]$, a supersymmetric generalisation of the bosonic symmetry algebra ${\rm hs}[\lambda]$;
the details of this symmetry algebra are spelled out in \cite{Gaberdiel:2013vva}, but are not very important 
for us in this paper.  One finds a spectrum of higher spin gauge 
fields which matches precisely with (\ref{D2mult2}). 
It was further shown in \cite{Gaberdiel:2014yla} that the asymptotic symmetry algebra of this higher spin theory 
is the same classical ${\cal W}$-algebra that arises in the 't~Hooft limit of the cosets. Here 
the 't~Hooft parameter  $\lambda$ of the CFT, see eq.~(\ref{tH}), is to be identified with the parameter $\lambda$ 
entering in the higher spin gauge algebra. 

In addition to the higher spin fields, there are matter fields which correspond to $(0; {\rm f})$ and its 
conjugate in the CFT (with multi particle states corresponding to the $(0; \Lambda)$ primaries
made from a finite number of boxes and anti-boxes). Thus 
the perturbative part of the Vasiliev theory is captured by the subsector of the CFT 
\cite{Gaberdiel:2011zw,Creutzig:2013tja,Candu:2013fta}
\be\label{pert0}
{\cal H}^{({\rm pert})} = \bigoplus_{\Lambda}\, (0;\Lambda) \otimes \overline{(0;\Lambda^\ast)} \ .
\ee
Note that this is a closed subsector of the CFT under OPE (in the large $N$, $k$ limit) but it is not modular invariant. 
One modular invariant completion of this sector is to consider the diagonal modular invariant. This necessitates 
adding in the contribution of all $(\Lambda_+; \Lambda_-)$ primaries. As in the purely bosonic and the
${\cal N}=2$ coset examples of holography, there are then many `light' states in the spectrum, see e.g., 
the analysis of \cite{Gaberdiel:2013cca}.  These, or equivalently, the $(\Lambda; 0)$  primaries are 
additional `non-perturbative' states which have some reflection in the Vasiliev theory (albeit in a non-unitary 
semi-classical limit) 
\cite{Castro:2011iw,Gaberdiel:2012ku,Tan:2012xi,Datta:2012km,Perlmutter:2012ds,Hikida:2012eu,%
Campoleoni:2013lma},
but whose precise role has not been completely elucidated.\footnote{Some of their properties
were analysed in \cite{Papadodimas:2011pf,Chang:2011vk,Jevicki:2013kma,Chang:2013izp}, 
see also \cite{Chang:2013izp} for an alternative interpretation.}  In Section~\ref{sec:comp} we will describe 
a different  modular invariant completion of (\ref{pert0})  in the $k \rightarrow \infty$ limit, which describes a 
symmetric  product CFT; for this modular invariant these light states (and many other primaries) are absent. 

\section{The Continuous Orbifold and the Symmetric Orbifold}\label{sec:3}

In this section we will make a preliminary identification of the large level limit of the Wolf space cosets with the 
symmetric  product orbifold. This will be based on the fact that this limit yields a continuous orbifold of free fermions 
and bosons. 
After explaining how this comes about, we will point out the close relation to the symmetric product orbifold.  

\subsection{The $k\rightarrow \infty$ Limit of the Cosets}\label{sec:cosetlim}

We are interested in the limit $k\rightarrow \infty$ of the above Wolf space cosets for which 
\be
\lambda=\gamma=0 \ , \qquad c = 6 (N+1) \ . 
\ee
In this limit, the large ${\cal N}=4$ superconformal algebra (\ref{A1}) -- (\ref{A8}) contracts to the small
superconformal ${\cal N}=4$ algebra together with $4$ free bosons and fermions \cite{Gukov:2004fh}, see also
Appendix~\ref{app:susy} for more details. One of the two $\mathfrak{su}(2)$ affine algebras becomes a global 
(custodial) symmetry in this limit and we are left with only one affine $\mathfrak{su}(2)$ algebra at level $(N+1)$, 
which is the R-symmetry algebra of the small ${\cal N}=4$ theory with $c=6(N+1)$. 

One can also understand rather easily what happens to the full coset spectrum in this limit. 
For large $k$, both the  ${\rm SU}(N+2)$ and the ${\rm SU}(N)$ group manifolds in the 
numerator and denominator decompactify, 
leading altogether to $4(N+1)$ free bosons. We also have $4(N+1)$ free fermions (that are described by 
the $\mathfrak{so}(4N+4)_1$ factor in the numerator) 
for any value of $k$. The zero modes of the denominator $\mathfrak{su}(N)\times \mathfrak{u}(1) \cong \mathfrak{u}(N)$ 
give rise to a gauging of these fermions and bosons (but without any kinetic term for the gauge fields). Since the 
$\mathfrak{u}(N)$ is embedded in the $\mathfrak{su}(N+2)$ in terms of an $N\times N$ 
diagonal block, we see that there are four free fermions and bosons which are uncharged with respect to the 
$\mathfrak{u}(N)$ --- these are precisely the $4$ free bosons and fermions that appear in the limit of
the large ${\cal N}=4$ superconformal algebra, as explained at the beginning of this section. The rest 
of the free bosons and fermions transform in the 
fundamental and anti-fundamental representation. Thus with respect to 
the $\mathfrak{u}(N)$
and the surviving $\mathfrak{su}(2)_{N+1}$ R-symmetry we have \cite{Gaberdiel:2013vva}
\begin{eqnarray}\label{bosfertrans0}
\hbox{bosons:} & \qquad & 2 \cdot ({\bf N},{\bf 1}) \oplus  2 \cdot (\overline{\bf N},{\bf 1})  \oplus 
4 \cdot ({\bf 1},{\bf 1}) \nonumber \\
\hbox{fermions:} & \qquad & ({\bf N},{\bf 2}) \oplus (\overline{\bf N},{\bf 2}) \oplus 2 \cdot ({\bf 1},{\bf 2})\  . 
\end{eqnarray}
We may therefore expect the limiting theory to be a supersymmetric continuous orbifold of the form\footnote{We will 
not be considering states that carry finite momentum or any other charge along the decompactified directions of the 
coset --- those would require us to consider primaries whose dimensions scale as $k$ as we take the contraction 
described in Appendix~\ref{app:susy}. We will therefore refer to this orbifold, in an abuse of notation, 
as $({\mathbb T}^4)^{N+1}/{\rm U}(N)$ though we are effectively considering it as 
$({\mathbb R}^4)^{N+1}/{\rm U}(N)$. This is appropriate since we will be comparing to the states of the 
corresponding  symmetric product  ${\rm Sym}_{N+1} (\mathbb{T}^4)$ without any
momentum/winding on the torus.} $({\mathbb T}^4)^{N+1}/{\rm U}(N)$.

In support of this identification, note that the coset primary $(0;{\rm f},(N+2))$ and its conjugate have, in this limit, 
conformal dimensions 
\be\label{ffer}
h(0;{\rm f},(N+2)) = h (0;\bar{\rm f},-(N+2)) = \frac{k+\frac{3}{2}}{2(N+k+2)} \cong \frac{1}{2} \ .
\ee
Thus we can, in a certain sense, think of $(0;{\rm f},(N+2))$ as corresponding to 
$2N$ free fermions transforming in the $({\bf N},{\bf 2})$ with respect to 
$\mathfrak{u}(N) \oplus \mathfrak{su}(2)$, and similarly for the conjugate representation. More
accurately, for example the ground state of the sector 
\be
(0;\bar{\rm f}) \otimes \overline{(0;{\rm f})} \quad \hbox{corresponds to} \quad 
\sum_{i}\, \psi^{* \bar{\imath}\alpha} \, \bar\psi^{i\beta}  \ , 
\ee
and similarly for the complex conjugate. (We denote here and in the following the right-moving
fields with a bar.) Furthermore, 
the free bosons are also contained in this description because they are the  
superconformal descendants of the free fermions, i.e., the $1/2$ descendants of the ground states,
see eq.~(2.40) of \cite{Gaberdiel:2013vva}. 

More generally, all the primaries $(0; \Lambda)$ (with a finite number of boxes and anti-boxes)
can be obtained from $(0;{\rm f})$ and $(0;\bar{{\rm f}})$ by repeated fusion, and they all have (half-)integral conformal 
dimension, as can be seen directly from (\ref{N4hdef})  --- in fact, for $k\rightarrow \infty$ only the $h=n$ term
survives. The 
${\rm U}(N)$ singlet condition (that applies simultaneously to left- and right-movers) then only guarantees that
the left- and right-moving states transform in conjugate coset representations, see the analogous discussion
in Section~2.2 of  \cite{Gaberdiel:2011aa}, as well as the more recent ${\cal N}=2$ analysis of \cite{GK}.
Thus the untwisted sector of the continuous orbifold
of the free theory is precisely accounted for by (\ref{pert0}), i.e., it matches exactly the `perturbative' 
spectrum of the dual higher spin theory.

We will later describe additional evidence, involving  the twisted sectors, that the limiting theory is indeed a 
continuous supersymmetric orbifold $({\mathbb T}^4)^{N+1}/{\rm U}(N)$ with the charges given in 
(\ref{bosfertrans0}). Orbifolding by a continuous group introduces twisted sectors with arbitrarily small twists -- 
and hence conformal dimensions. From the point of view of free fermions and bosons which are charged 
in the fundamental and anti-fundamental of ${\rm U}(N)$ these are states with arbitrarily small gauge 
holonomy. These give rise to the 
light states mentioned in the previous section. More general twists lead to the other primaries 
$(\Lambda_+; \Lambda_-)$. In fact, as mentioned earlier, the diagonal modular invariant partition function 
necessarily contains all these twisted sector states. We will be considering in the next section, a 
non-diagonal modular invariant combination in which all the light states are automatically absent.

\subsection{Relation to the Symmetric Product CFT}\label{sec:3.2}

We now want to relate the continuous orbifold by ${\rm U}(N)$ to a discrete orbifold by $S_{N+1}$. 
Let us first focus on the untwisted sector of the continuous orbifold. We want to show that it is a natural
subsector of the untwisted sector of the symmetric orbifold. 
In fact, given the explicit description of the continuous orbifold from above, this is now almost immediate. 
Recall that, in the symmetric orbifold
\be
{\rm Sym}_{N+1} (\mathbb{T}^4)  \equiv \left. \bigl( \mathbb{T}^4 \bigr)^{N+1}  \right/ S_{N+1} \ , 
\ee
the untwisted sector of this theory\footnote{As mentioned in Section~\ref{sec:cosetlim}, we shall only consider 
the subspace of  states that do not carry any momentum along the $\mathbb{T}^4$.} consists of
 $4(N+1)$ free bosons and fermions that transform as 
\begin{eqnarray}\label{bosfertranssym0}
\hbox{bosons:} & \qquad & 4 \cdot ( N+1,{\bf 1}) \nonumber \\
\hbox{fermions:} & \qquad & 2\cdot (N+1,{\bf 2}) 
\end{eqnarray}
with respect to $S_{N+1} \times \mathfrak{su}(2)$, where the $\mathfrak{su}(2)$ is the 
R-symmetry of the small ${\cal N}=4$ superconformal algebra.

Next we recall that the `defining' $(N+1)$-dimensional representation of $S_{N+1}$ 
(in terms of permutation matrices) is not irreducible. In fact, it always contains a $1$-dimensional
subspace corresponding to the sum of all the $N+1$ basis vectors; thus, as a representation
of $S_{N+1}$, we have the decomposition
\be\label{symdec}
N+1 =  1 \oplus  N \ , 
\ee
where $N$ denotes the so-called `standard' representation of $S_{N+1}$. (It corresponds to the Young
diagram consisting of $N$ boxes in the first row, and one box in the second.)
We note that this decomposition also gives rise to a natural embedding of 
\be\label{subgroup}
S_{N+1} \subset {\rm U}(N) \ ,
\ee
see Appendix~\ref{app:embedding} for an explicit description.
Indeed, to every $(N+1)\times (N+1)$ permutation matrix, we can associate the unitary 
$N\times N$ matrix that acts on the $N$-dimensional subspace in (\ref{symdec}). We can
therefore ask how the various representations of ${\rm U}(N)$ decompose with respect to 
$S_{N+1}$, and one finds that we have the branching rules
\be\label{branchrule}
{\bf N}^{{\rm U}(N)} \ \rightarrow \ N^{S_{N+1}} \ , \qquad
\overline{\bf N}^{{\rm U}(N)} \ \rightarrow \ N^{S_{N+1}} \ , 
\ee
where in both cases the $S_{N+1}$ representation is the standard representation from 
above.\footnote{We have explicitly displayed the superscripts to indicate the relevant group 
but we shall drop these with the understanding that boldface represents ${\rm U}(N)$ and roman face 
represents $S_{N+1}$.} 
Here it is important that we are dealing with ${\rm U}(N)$ rather than ${\rm SU}(N)$. Then
$\overline{\bf N}$ is the complex conjugate representation of ${\bf N}$ 
(with opposite ${\rm U}(1)$ charge); since the representations of $S_{N+1}$ are all real,
the branching of ${\bf N}$ and $\overline{\bf N}$ with respect to (\ref{subgroup}) are therefore the same.

These observations now have an important consequence. As we have explained above, the 
untwisted sector of the continuous orbifold can be described in terms of $4(N+1)$ free bosons
and fermions that transform as in eq.~(\ref{bosfertrans0}) with respect to ${\rm U}(N) \times \mathfrak{su}(2)$. 
If we consider these states with respect to the subgroup $S_{N+1} \times \mathfrak{su}(2)$, it then
follows from eq.~(\ref{branchrule}) that they transform simply as 
\begin{eqnarray}\label{bosfertranssym}
\hbox{bosons:} & \qquad & 4 \cdot (N,{\bf 1})  \oplus 4 \cdot  ( 1,{\bf 1})  \nonumber \\
\hbox{fermions:} & \qquad & 2\cdot (N,{\bf 2})  \oplus 2\cdot (1,{\bf 2})  \ , 
\end{eqnarray}
i.e., precisely as (\ref{bosfertranssym0}). Thus the symmetric orbifold action is induced by the embedding
of $S_{N+1}\subset {\rm U}(N)$. 

We therefore conclude that the ${\rm U}(N)$ singlet sector of this free theory (i.e., eq.~(\ref{pert0}))
is a subsector of the $S_{N+1}$ singlet sector, i.e., of the untwisted sector of the symmetric orbifold. 
Since the former captures precisely the perturbative higher spin degrees of freedom, this is a very 
precise CFT incarnation of the belief that, at the tensionless point in moduli space, the higher spin theory 
should form a consistent subsector of string theory! 

More generally, the partition function of the symmetric orbifold can be viewed as a non-diagonal modular 
invariant of the continuous orbifold theory. In order to understand this, let us consider a general
pair of orbifold CFTs 
\be\label{orbgen}
{\cal H}_1  = {\cal H}^{(0)}/G \qquad \hbox{and} \qquad {\cal H}_2= {\cal H}^{(0)}/H \ , \qquad 
\hbox{where $H \subset G$,}
\ee
and  the action of $H$ in the orbifold ${\cal H}_2$ is induced from that of $G$ in ${\cal H}_1$. 
Then the untwisted sector of ${\cal H}_1$ will be {\em contained} in the untwisted sector of ${\cal H}_2$ --- only 
those states in the untwisted sector of ${\cal H}_2$ that are also invariant with respect to the whole action of $G$ 
will survive in the untwisted sector of ${\cal H}_1$. In particular, we can therefore organise ${\cal H}_2$
in terms of representations of the chiral algebra of ${\cal H}_1$, i.e., in terms of the $G$-invariant
generators of the original chiral algebra. Thus we can think of ${\cal H}_2$ as a non-diagonal 
modular invariant of the chiral algebra of ${\cal H}_1$. We will see an explicit realisation of this idea
(with $H=S_{N+1}\subset {\rm U}(N)=G$) in the following section.

\section{Comparison to the Symmetric Orbifold Theory}\label{sec:comp}

The schematic structure of the partition function of the symmetric orbifold is (in the NS sector) 
\be\label{symmpart}
Z_{\rm NS}(q,\bar{q}, y, \bar{y})= |{\cal Z}_{\rm vac}(q,y)|^2+ \sum_{j}|{\cal Z}^{(\rm U)}_{j}(q,y)|^2
+ \sum_{\beta,l}|{\cal Z}^{(\rm T)}_{\beta,l}(q,y)|^2 \ .
\ee
Here ${\cal Z}_{\rm vac}$ is the vacuum character of the symmetric product CFT, $j$ 
labels the nontrivial primaries in the untwisted sector (built from the orbifold invariant combinations 
of free fermions and their descendants that are not just products of chiral and anti-chiral fields), and 
${\cal Z}^{(\rm T)}_{\beta,l}$ are the corresponding contributions from the twisted sectors (where $\beta$ labels 
the different non-trivial conjugacy classes of $S_{N+1}$). The above considerations imply that we should 
be able to reorganise each of these in terms of characters of the continuous orbifold; from 
this viewpoint, the above partition function then corresponds to a 
non-diagonal modular invariant of the continuous orbifold algebra. In this section, we illustrate this for the 
vacuum sector as well as the simplest nontrivial primary of the untwisted sector. 
In the next section, we generalise this to the twisted sector. 

\subsection{The Chiral Algebra of the Symmetric Orbifold}\label{sec:Wext}

We will now argue that we can write 
\be\label{vacgen}
{\cal Z}_{\rm vac}(q,y) 
= \sum_{\Lambda} n(\Lambda) \, \chi_{(0;\Lambda)}(q,y) \ ,
\ee 
where $n(\Lambda)$ are positive integers. 
Here the characters on the RHS are the coset characters in the limit as $k\rightarrow \infty$. 
By the arguments at the end of the previous section, we should be able to write the vacuum representation 
of the symmetric orbifold theory as such a linear combination since the untwisted sector of the 
${\rm U}(N)$ orbifold comprises the representations $(0;\Lambda)$. 
The integer multiplicities $n(\Lambda)$ consequently have a simple interpretation: they are the 
multiplicity with which the trivial representation of $S_{N+1}$ appears in the corresponding 
${\rm U}(N)$ representation $\Lambda$ under the branching (\ref{subgroup}).
An immediate consequence of (\ref{vacgen}) is that the ${\cal W}$-algebra of the symmetric orbifold
theory should be a huge (infinite) extension of the Wolf space coset ${\cal W}$-algebra in the limit of 
$k\rightarrow \infty$, i.e., of ${\cal W}_{\infty}[0]$. The multiplicities $n(\Lambda)$ can be worked out as 
explained in  Appendix~\ref{app:singlets}.

This abstract reasoning can now be tested concretely. We can compute independently the LHS and RHS of 
(\ref{vacgen}) using various known facts about the symmetric product CFT and the coset CFT, respectively. 
Let us first focus on the LHS. 

The generating function for the untwisted sector in the R-R sector of the symmetric orbifold is 
\cite{Dijkgraaf:1996xw}
\be\label{generating}
\sum_{k=0}^{\infty} p^k Z^{({\rm U})} ({\rm Sym}^k(X)) = \prod_{\Delta,\bar\Delta,\ell,\bar\ell} 
\frac{1}{(1 - \, p q^{\Delta} \bar{q}^{\bar\Delta} y^{\ell} \bar{y}^{\bar{\ell}} )^{ 
c(\Delta,\bar\Delta, \ell,\bar\ell) }} \ , 
\ee
where the coefficients $c(\Delta,\bar\Delta,\ell,\bar\ell)$ are the expansion coefficients of the
R-R partition function of $X$ (with the insertion of $(-1)^{F+\tilde{F}}$),
\be
Z(X) = \sum_{\Delta,\bar\Delta,\ell,\bar\ell}   c(\Delta,\bar\Delta, \ell,\bar\ell) \, 
q^{\Delta} \bar{q}^{\bar\Delta} y^{\ell} \bar{y}^{\bar{\ell}} \ .
\ee
We are interested in the ${\cal W}$ algebra of this theory, i.e., we want to analyse only
the purely left-moving states and we want to describe them in the NS-sector.

For the case at hand, $X=\mathbb{T}^4$, the partition function of $X$ factorises into left- and
right-movers, whose chiral part equals
\be\label{chiralT}
Z_{\rm chiral}(\mathbb{T}^4) = - \Bigl( \frac{\vartheta_1(z|\tau)}{\eta(\tau)}\Bigr)^2 \, \frac{1}{\eta^4(\tau)} \ , 
\ee
where 
\be
\vartheta_1(z|\tau) = i (y^{1/2} - y^{-1/2}) \, q^{\frac{1}{8}} \, \prod_{n=1}^{\infty} 
(1-q^n)\, (1-y q^n) (1 - y^{-1} q^n) \ . 
\ee
Because of this factorisation property, we can simply work out the ${\cal W}$-algebra by 
spectrally flowing the chiral version of (\ref{generating}) to the NS sector. 
In order to do so explicitly, we need the first few expansion coefficients of (\ref{chiralT}) 
\begin{eqnarray}
Z_{\rm chiral}(\mathbb{T}^4) 
& = &  \bigl(y - 2 + y^{-1}\bigr) 
+  \bigl( - 2 y^2 + 8 y - 12 + 8 y^{-1} - 2 y^{-2} \bigr)  \, q  \nonumber \\
& & {}+ \bigl( y^3 - 12 y^2 + 39 y - 56 + 39y^{-1} - 12y^{-2}+1y^{-3} \bigr) \, q^2  \nonumber \\
& & {}+ \bigl( 8y^3 - 56 y^2+152 y - 208+152 y^{-1} - 56y^{-2}+8 y^{-3}  \bigr) \, q^3 \nonumber \\
& & {}+ \bigl( - 2 y^4+39 y^3 -208 y^2+513 y - 684 +513y^{-1} - 208 y^{-2} \nonumber \\
& & \qquad + 39y^{-3} - 2 y^{-4} \bigr) \, q^4 \nonumber \\
& & {}+ {\cal O}(q^5) \ .   \label{Zexp}
\end{eqnarray}
Then we plug the corresponding coefficients $c(\Delta,\ell)$ into the chiral version of the generating
function (\ref{generating}), and flow to the NS-sector (without the insertion of $(-1)^F$) by replacing 
\be
y \mapsto - y \, q^{\frac{1}{2}} \ , \qquad 
p \mapsto - p \, q^{\frac{1}{4}} \,y   \ . 
\ee
The first few terms of the ${\cal W}$-algebra character equal (we are considering here the case
where we have taken sufficiently many copies, i.e., consider a sufficiently high power of $p$, so that the coefficients 
stabilise, and we ignore the overall factor of $q^{-c/24}$)
\begin{eqnarray}
{\cal Z}_{\rm vac}(q,y) & =  & 
1 + \bigl( 2 y+2 y^{-1} \bigr) q^{\frac{1}{2}} 
+ \bigl(2 y^2 + 12+2 y^{-2}\bigr) q \nonumber \\
& & {}+ \bigl(2 y^3+32 y+32 y^{-1}+2 y^{-3} \bigr) q^{\frac{3}{2}}
\nonumber \\
& & {}+ \bigl( 2 y^4+52 y^2+159+52 y^{-2}+2 y^{-4} \bigr) \, q^2 \nonumber \\
& & {}+ \bigl(2 y^5+62 y^3+426 y+426 y^{-1}+62y^{-3}+2 y^{-5} \bigr)\, q^{\frac{5}{2}} \label{Wsym} \\
& & {}+ \bigl( 2 y^6 + 64 y^4 + 767 y^2 + 1800 + 767 y^{-2} + 64 y^{-4} + 2 y^{-6} \bigr)\, q^3 + {\cal }O(q^{\frac{7}{2}}) \ . \nonumber
\end{eqnarray}

We can now compare this with the RHS of eq.~(\ref{vacgen}). To compute the latter to the order $q^3$, we 
need the multiplicities $n(\Lambda)$ for representations $\Lambda$ with up to six boxes and anti-boxes.
These are given at the end of Appendix~\ref{app:singlets}. The corresponding coset characters
$\chi_{(0;\Lambda)}(q,y)$, with nonzero multiplicities, (and at large $N$ with $k\rightarrow\infty$) can be 
computed following the techniques developed  in \cite{Candu:2012jq,Candu:2013fta}; the details are explained in 
Appendix~\ref{app:coset}. One finds  remarkable agreement, at least to the order we have 
computed,\footnote{In (\ref{Wcharexp}) we have written out 
all representations with $B\leq 4$, but only those representations with $B=5,6$ that have a non-trivial 
contribution at order $q^3$.} namely
\begin{eqnarray}\label{Wcharexp}
{\cal Z}_{\rm vac}(q,y)  & = &  \chi_{(0;0)}(q,y)  + 
\chi_{(0;[2,0,...,0])}(q,y) + \chi_{(0;[0,0,...,0,2])}(q,y) \nonumber \\
& & + \ \chi_{(0;[3,0,...,0,0])}(q,y) + \chi_{(0;[0,0,0,...,0,3])}(q,y) \nonumber \\
& & +\  \chi_{(0;[2,0,...,0,1])}(q,y) + \chi_{(0;[1,0,0,...,0,2])}(q,y) \nonumber \\
& & +  \ 2 \cdot \chi_{(0;[4,0,...,0,0])}(q,y) + 2 \cdot \chi_{(0;[0,0,0,...,0,4])}(q,y) \nonumber  \\
& & + \ \chi_{(0;[0,2,0,...0,0])}(q,y) +  \chi_{(0;[0,0,...0,2,0])}(q,y) \nonumber \\
& & + \ \chi_{(0;[3,0,...,0,1])}(q,y) + \chi_{(0;[1,0,0,...,0,3])}(q,y) \nonumber \\
& & + \ 2\cdot \chi_{(0;[2,0,0,...,0,2])}(q,y) \nonumber  \\
& & + \ \chi_{(0;[1,2,0,...,0])}(q,y) + \chi_{(0;[0,...,0,2,1])}(q,y) \nonumber \\
& & + \ \chi_{(0;[2,1,0,...,0,1])}(q,y) + \chi_{(0;[1,0,...,0,1,2])}(q,y) \nonumber \\
& & + \ \chi_{(0;[0,2,0,...,0,1])}(q,y) + \chi_{(0;[1,0,...,0,2,0])}(q,y) \nonumber \\
& & + \ 3\cdot \chi_{(0;[3,0,...,0,2])}(q,y) + 3\cdot \chi_{(0;[2,0,...,0,3])}(q,y) \nonumber \\
& & + \  \chi_{(0;[1,1,0,...,0,2])}(q,y) +  \chi_{(0;[2,0,...,0,1,1])}(q,y)  \nonumber \\
& & + \  \chi_{(0;[0,0,2,0,...,0])}(q,y) +  \chi_{(0;[0,...,0,2,0,0])}(q,y) \nonumber \\
& & + \ 3\cdot \chi_{(0;[0,2,0,...,0,2])}(q,y) + 3\cdot \chi_{(0;[2,0,...,0,2,0])}(q,y) \nonumber \\
& & + \  \chi_{(0;[1,1,0,...,0,1,1])}(q,y) + \ {\cal O}(q^{7/2}) \ . 
\end{eqnarray}
This is nontrivial evidence for the correctness of (\ref{vacgen}) and therefore for the arguments that led to 
it.  Presumably one should be able to prove analytically 
the mathematical identity expressed by (\ref{vacgen}), but we have not tried to do so.

\subsection{Microscopic Realisation}\label{4.2.}

Before we come to describing how some of the other states of the symmetric orbifold 
can be organised in terms of ${\cal W}_{\infty}[0]$ representations, let us pause
for a moment and understand the structure of (\ref{Wcharexp}) directly in terms of the free fermions
and bosons that appear in the coset description for $k\rightarrow \infty$. Let us denote the
(left-moving)  fermions
that transform as $({\bf N}\oplus {\bf 1},{\bf 2})\oplus (\overline{\bf N} \oplus {\bf 1},{\bf 2})$ by $\psi^{i\alpha}$ and 
$\psi^{* \bar{\jmath}\beta}$, respectively, while the bosons that transform in the 
$2\cdot ({\bf N}\oplus{\bf 1},{\bf 1}) \oplus 2\cdot (\overline{\bf N}\oplus{\bf 1},{\bf 1})$ 
are  $\phi^{ia}$ and $\phi^{*\bar{\jmath}b}$,
respectively. Here $i,j=1,\ldots,N+1$, while $\alpha,\beta\in\{-\frac{1}{2},\frac{1}{2}\}$ and
$a,b\in\{1,2\}$. The ${\cal W}_{\infty}$ algebra of the coset is generated by the bilinear 
combinations of these fields that are ${\rm U}(N)$ singlets. The additional generators of the 
symmetric orbifold chiral algebra, on the other
hand, are just invariant under the symmetric group. The simplest states that are invariant under
the symmetric group (but not under ${\rm U}(N)$) are
\be \label{new2}
\sum_{i=1}^{N+1} \psi^{i\alpha}_{-1/2} \psi^{i\beta}_{-1/2} |0\rangle \ , \qquad
\hbox{and} \qquad
\sum_{i=1}^{N+1} \psi^{*\bar{\imath}\alpha}_{-1/2} \psi^{*\bar{\imath}\beta}_{-1/2} |0\rangle \ .
\ee
Since the fermions satisfy anti-commutation relations, the only terms that contribute are 
those with $\alpha=-\beta$; thus the resulting states transform as singlets under
the R-symmetry $\mathfrak{su}(2)$, and they have conformal dimension $h=1$. They
can therefore be identified with the ground states of the coset representations
\be\label{20p02}
(0;[2,0,\ldots,0,0]) \qquad \hbox{and} \qquad
(0;[0,0,\ldots,0,2]) 
\ee
respectively, see eq.~(\ref{2000}). To be clear, there are also the states
\be\label{old2}
\sum_{i,j=1}^{N+1} \psi^{i\alpha}_{-1/2} \psi^{j\beta}_{-1/2} |0\rangle \ , \qquad
\hbox{and} \qquad
\sum_{i,j=1}^{N+1} \psi^{*\bar{\imath}\alpha}_{-1/2} \psi^{*\bar{\jmath}\beta}_{-1/2} |0\rangle \ ,
\ee
that are singlets with respect to the symmetric group. However, they are already contained in 
the ${\cal W}_{\infty}$ algebra itself since $\sum_i \psi^{i\alpha}$ and 
$\sum_i \psi^{*\bar{\imath}\alpha}$ correspond to the four free fermions of the large ${\cal N}=4$
superconformal algebra. Thus, strictly speaking, (\ref{20p02}) corresponds to a certain
linear combination of (\ref{new2}) and (\ref{old2}), which then transforms indeed in
the $[2,0,\ldots,0,0]$ and $[0,0,\ldots,0,2]$ of ${\rm U}(N)$, respectively.
\smallskip

In the same spirit (i.e., removing the analogues of (\ref{old2})), the 
states corresponding to $(0;[3,0,\ldots,0])$ can then be identified with 
\be
\sum_{i=1}^{N+1} \psi^{i\alpha}_{-1/2} \psi^{i\beta}_{-1/2} \phi^{ia}_{-1} |0\rangle \ , 
\ee
where because of the anti-symmetry of the fermions again $\alpha=-\beta$ and 
the third generator has to be a boson (for the state of lowest conformal dimension). 
Since $a$ takes two values, the leading contribution 
of these states is $2 q^2$, in agreement with the character of eq.~(\ref{3000}). On the other 
hand, the ground state of  $(0;[2,0,\ldots,0,1])$ is described by 
\be
\sum_{i=1}^{N+1} \psi^{i\alpha}_{-1/2} \psi^{i\beta}_{-1/2} \psi^{*\bar{\imath}\gamma}_{-1/2} |0\rangle \ , 
\ee
which transforms as a doublet under the R-symmetry $\mathfrak{su}(2)$ and has
$h=\frac{3}{2}$, again in agreement with its character. (The corresponding wedge character
is just the product of eq.~(\ref{1000}) and eq.~(\ref{2000}).) Something more interesting happens
for the case of $(0;[4,0,\ldots,0])$, for which we have 
\be\label{newp4}
\sum_{i=1}^{N+1} \psi^{i\alpha}_{-1/2} \psi^{i\beta}_{-1/2} \phi^{ia}_{-1} \phi^{ib}_{-1} |0\rangle  \qquad
\hbox{and} \qquad
\sum_{i,j=1}^{N+1} \psi^{i\alpha}_{-1/2} \psi^{i\beta}_{-1/2} \phi^{ja}_{-1} \phi^{jb}_{-1} |0\rangle \ . 
\ee
These states are singlets under the R-symmetry $\mathfrak{su}(2)$, and because they are 
symmetric under the exchange of the two bosonic generators give rise to $3$ states at $h=3$; 
thus each of them contributes $3 q^3$ to the character, which is indeed the leading term of the 
wedge character of $(0;[4,0,\ldots,0])$ that is given explicitly in the ancillary file. This therefore
explains the multiplicity of $2$ with which this representation contributes to (\ref{Wcharexp}).

We should note that there is also the closely related state 
\be\label{newp41}
\sum_{i,j=1}^{N+1} \psi^{i\alpha}_{-1/2} \psi^{j\beta}_{-1/2} \phi^{ia}_{-1} \phi^{jb}_{-1} |0\rangle \ ,
\ee
which however has somewhat different symmetry properties. This state (or rather a linear combination
of this state and the second state in (\ref{newp4})) appears as  a descendant of the four fermion term
\be\label{new4}
\sum_{i,j=1}^{N+1} \psi^{i\alpha}_{-1/2} \psi^{j\beta}_{-1/2} \psi^{i\gamma}_{-1/2} \psi^{j\delta}_{-1/2} |0\rangle  \ ,
\ee
which corresponds to the leading term in $(0;[0,2,0,\ldots,0])$ --- indeed in (\ref{new4}) the anti-symmetry
of the fermions implies that there is a single singlet state under the R-symmetry $\mathfrak{su}(2)$, in
agreement with the leading term of the corresponding wedge character that is also given in the ancillary file. 

The higher terms can be constructed similarly,
although this becomes more and more cumbersome. (In effect, this construction is just an 
explicit incarnation of the counting argument of Appendix~\ref{app:singlets}.)

\subsection{Other Representations in the Untwisted Sector}\label{4.3}

After this brief interlude let us return to describing the structure of the untwisted sector of the symmetric orbifold. 
As mentioned before, it does not just contain the contribution from
the extended ${\cal W}_{\infty}[0]$ vacuum character, i.e., the mod squared of the character 
(\ref{Wcharexp}). Additional representations of this algebra, whose characters are denoted by 
${\cal Z}^{(\rm U)}_{j}$ in (\ref{symmpart}), also appear. They can, in turn, be written as direct sums of 
${\cal W}_{\infty}[0]$ representations. Let us illustrate this for the simplest  non-trivial case. The untwisted 
sector of the symmetric  orbifold contains also the states of the form
\be
\sum_{i=1}^{N+1} \, \psi^{i\alpha}_{-1/2}\,  \bar\psi^{i\beta}_{-1/2} |0\rangle  \ , 
\ee
and similar combinations where either the left-moving $\psi^{i\alpha}$ or the right-moving $\bar\psi^{i\beta}$ (or both)
are replaced by the complex conjugate fermions. From the coset point of view, the corresponding
states are the ground states of 
\be\label{ferbi}
(0;{\rm f}) \otimes \overline{(0;{\rm f})} \ , \qquad 
(0;\bar{\rm f}) \otimes \overline{(0;{\rm f})} \ , \qquad 
(0;{\rm f}) \otimes \overline{(0;\bar{\rm f})} \ , \qquad 
(0;\bar{\rm f}) \otimes \overline{(0;\bar{\rm f})} \ , 
\ee
where the second factor (that is overlined) refers to the right-movers. Note that both $(0;{\rm f})$ and 
$(0;\bar{\rm f})$ are 
in the same representation of the extended ${\cal W}_{\infty}$ algebra since
\be
(0;{\rm f}) \otimes (0;[0,0,\ldots,0,2]) \supset (0;\bar{\rm f}) \ . 
\ee
Thus the contribution from all states in (\ref{ferbi}) will be part of the modulus square of a 
single extended  ${\cal W}_\infty$ representation. 

Using similar techniques as in the derivation of (\ref{Wsym}) we can determine the character
of this  extended ${\cal W}_\infty$ representation: in expanding out (\ref{generating}), we
now have to consider the multiplicities (as functions of $q$ and $y$) of the term 
$\bar{q}^{1/2} (2 \bar{y} + 2 \bar{y}^{-1})$, and subtract out the contribution coming 
from the mod square of (\ref{Wsym}). To do so, we consider the NS-sector version of (\ref{generating}), and 
expand out one factor in the denominator with $\bar\Delta = \frac{1}{2}$ to first order ---  the coefficients 
of  $\bar{q}^{1/2} (2 \bar{y} + 2 \bar{y}^{-1})$ from this factor are then just 
$Z^{({\rm NS})}_{\rm chiral}(\mathbb{T}^4)(q,y)$,  the NS-sector version of (\ref{Zexp}), see eq.~(\ref{E1}) 
for an explicit formula. For all the other factors from the denominator we take the term with 
$\bar\Delta=0$. After subtracting out the contribution from the mod squared of the extended
vacuum sector, this then leads to
\be\label{chi1}
{\cal Z}_1(q,y) =  {\cal Z}_{\rm vac}(q,y) \, \bigl[ Z_{\rm chiral}^{({\rm NS})} (\mathbb{T}^4) (q,y) - 1 \bigr] \ .
\ee
The resulting expression for ${\cal Z}_1(q,y)$ is then given 
explicitly (to low order) in eq.~(\ref{E2}). 
Just as for the vacuum character, we can argue on general grounds that 
\be\label{fundgen}
{\cal Z}_1(q,y) 
= \sum_{\Lambda} n_1(\Lambda) \, \chi_{(0;\Lambda)}(q,y) \ ,
\ee 
where $n_1(\Lambda)$ is the multiplicity with which the standard $N$ dimensional representation of 
$S_{N+1}$ appears in the ${\rm U}(N)$ representation $\Lambda$. Again these multiplicities can be worked 
out using the techniques of Appendix~\ref{app:singlets}. 

In Appendix~\ref{app:untwist}, we explicitly verify 
this identity by expanding out both sides of eq.~(\ref{fundgen}) to order $q^{3}$ --- see eq.~(\ref{Wcharexp2}). 
Note that $|{\cal Z}_1(q,y)|^2$ accounts (among others) for the states 
\be
\sum_{i=1}^{N+1} \, \phi^{ia}_{-1}\,  \bar\phi^{ib}_{-1} |0\rangle  \ , 
\ee
as well as the combinations for which either $\phi^{ia}$ or $\bar\phi^{ib}$ (or both) are replaced
by the corresponding complex conjugate operators. Altogether these states are the $16 = 4 \times 4$
exactly marginal operators that preserve the small ${\cal N}=4$ superconformal algebra and
deform the shape (and complex structure) of the $\mathbb{T}^4$. 

\section{Twisted Sectors and Light States}\label{sec:light}

As we have seen above, the untwisted sector of the symmetric orbifold can be written
in terms of representations of  the ${\cal W}_{\infty}[0]$ algebra of the Wolf space cosets.
Indeed, the chiral algebra of the symmetric orbifold is a certain extension of the  
${\cal W}_{\infty}[0]$ algebra, see eq.~(\ref{Wcharexp}), and similar statements apply
to the other representations that appear in the untwisted sector of the symmetric orbifold, see
e.g., the discussion of eq.~(\ref{fundgen}) above. In this section we want to discuss qualitatively the structure
of the twisted sector of the symmetric orbifold; a more quantitative analysis is the topic of the 
subsequent two sections.

In order to understand what we should expect, let us go back to the generic example of 
two orbifolds as in eq.~(\ref{orbgen}). As we have explained there, the untwisted sector of 
${\cal H}_1$ is contained in the untwisted sector of ${\cal H}_2$, since invariance under 
$G$ imposes more constraints than invariance under the subgroup $H\subset G$. What can 
we say about the twisted sectors? As a first rough guide, modular invariance essentially implies
that the number of states is determined by the central charge, and hence does not change upon 
orbifolding. Thus if ${\cal H}_1$ has {\em fewer} states than ${\cal H}_2$ in the untwisted
sector, it will contain {\em more} states in the twisted sector (so that the total number is roughly the same). 
More concretely, for any orbifold, the twisted sectors are labelled by the conjugacy classes of the orbifold group,
and the conjugacy classes of a subgroup $H\subset G$ are naturally contained in the 
conjugacy classes of $G$.\footnote{Note though that this `embedding' does not need to be injective, i.e.,
different conjugacy classes of $H$ may map to the same conjugacy class of $G$.} Thus the twisted
sectors of the orbifold ${\cal H}_2$ will be contained in the twisted sectors of ${\cal H}_1$.

For the case at hand where $G={\rm U}(N)$, the conjugacy classes of $G$ are para\-me\-trised 
by the Cartan torus ${\rm U}(1)^N$ modulo the action of the Weyl group $S_N$; on the other hand, the 
conjugacy classes of 
$H=S_{N+1}$ are finite in number, and are labelled by the partitions of $N+1$. For any group element in 
$S_{N+1}$ we can determine the $N$ eigenvalues in the standard representation of $S_{N+1}$, 
and these eigenvalues are (up to permutation) the same for each representative of a given 
conjugacy class  of $S_{N+1}$. Thus we can naturally associate to every conjugacy class of $S_{N+1}$ an
element in ${\rm U}(1)^N / S_N$, i.e., a conjugacy class in $G={\rm U}(N)$.

These arguments therefore imply that the twisted sectors of the symmetric product orbifold are 
a subset of the twisted sectors of the continuous orbifold. This phenomenon also has a natural
interpretation from the viewpoint of the representation theory of the continuous orbifold: every
representation of the symmetric orbifold chiral algebra is obviously also a representation
of the continuous orbifold chiral algebra (since the former chiral algebra is an extension of the latter, see
eq.~(\ref{Wcharexp})), but the converse is in general not true: a representation of the continuous
orbifold chiral algebra does not necessarily define a representation of the extended chiral algebra
of the symmetric orbifold. Indeed, 
a necessary condition is that the additional primaries that appear in eq.~(\ref{Wcharexp})
must be local with respect to the representation at hand, and this is not automatic. This is 
the representation theoretic reason why not all twisted sectors of the continuous orbifold
will give rise to (local) representations of the symmetric orbifold, i.e., why they do not appear 
in the spectrum.  

Unfortunately, this locality condition is in general
quite hard to analyse, see however Section~\ref{sec:local}, 
and we cannot easily determine which representations of the Wolf space coset actually
give rise to representations of the symmetric orbifold. However, we can show that the expected representations
of the symmetric orbifold indeed arise in this manner, and we will give an example of that below,
see Section~\ref{2cycle}.

\subsection{Light States and Quantisation of the Higher Spin Theory}\label{5.1}

When we extend the chiral algebra from ${\cal W}_{\infty}[0]$ to that
of the symmetric orbifold only those twisted sector representations of the continuous orbifold become
representations of the extended chiral algebra that correspond to the discrete twists in 
$S_{N+1}\subset {\rm U}(N)$. Note that, in particular, all the `small' twists in ${\rm U}(N)$ that correspond
to the so-called `light states' of the cosets do {\em not} survive --- thus by embedding the higher spin 
theory into string theory, the `non-perturbative' states of the higher spin theory that are dual to the light states 
\cite{Castro:2011iw} do not lift to solutions of string theory. In particular, these `light states' are therefore
absent in string theory. 

Thus when we try to quantise the higher spin theory, then on the level of
the dual CFT there are at least two natural choices. Either we consider the standard charge conjugation modular
invariant of the dual CFT --- then we do not add any perturbative degrees of freedom to the 
higher spin theory, but the consistency of the CFT (in particular modular invariance) requires us to
add many low-lying non-perturbative solutions of the higher spin theory, i.e., the duals of the light states. 

The other alternative is that we extend the chiral algebra of the dual CFT by embedding it into the
chiral algebra of some string theory, i.e., in our case, the chiral algebra of the symmetric orbifold. Then
this corresponds to adding {\em many perturbative degrees of freedom} to the higher spin theory. 
However then the consistency of the dual CFT does not require to add any further low-lying 
degrees of freedom, i.e., the light states do not appear in the spectrum any longer. 

In either case, we see that the quantisation of the higher spin theory requires us to add low-lying
degrees of freedom --- either the light states or the perturbative string degrees of freedom. This
suggests that a direct quantisation of the higher spin theory by itself is problematic.

\section{The Twisted Sector of the Continuous Orbifold}\label{sec:6}

Let us now try to make the statements of the previous section more concrete. In order to do so,
we first need to understand how the twisted sector representations of the continuous orbifold
can be described in terms of the coset language. Unfortunately, the following discussion is slightly
technical; readers who are not interested in the detailed
derivation and justification of the correspondence, see eq.~(\ref{twist}), may skip this section
and jump directly to Section~\ref{sec:7}.

\subsection{The Twisted Sector Ground States}\label{sec:twisted}

As we have explained above, the twisted sectors of the continuous orbifold 
are labelled by elements of the Cartan torus, modulo the Weyl group. For the case at hand,
the Cartan torus of ${\rm U}(N)$ is simply ${\rm U}(1)^N$, and the Weyl group is the symmetric group
that permutes the $N$ ${\rm U}(1)$ factors. Thus the twisted sectors are labelled by $N$-tuples
\be
[\alpha_1,\ldots,\alpha_N] \qquad \tfrac{1}{2} \geq \alpha_1 \geq \alpha_2 \geq \cdots \geq \alpha_N \geq -\tfrac{1}{2} \ .
\ee
Given that each $\alpha_i$ describes the `twist' of $4$ bosons and fermions, the conformal dimension of the 
corresponding twisted sector ground state should then equal
\be\label{ground}
h \bigl([\alpha_1,\ldots,\alpha_N] \bigr) = \sum_{i=1}^{N} |\alpha_i | \ . 
\ee
Indeed, each $\alpha$-twisted complex boson and fermion contributes
\be
\Delta h_{\rm bos} = \frac{1}{2} \alpha(1-\alpha) \ , \qquad \Delta h_{\rm fer} = \frac{1}{2} \alpha^2 \ , 
\ee
where the bosonic formula holds for $0\leq \alpha \leq 1$, while the fermionic formula is correct for 
$0\leq |\alpha|\leq \frac{1}{2}$. The total ground state energy of a twisted complex boson-fermion pair is then
$\frac{1}{2} |\alpha|$, and since $4$ bosons and fermions correspond to two such pairs, we get altogether
(\ref{ground}). 

Unlike the bosonic situation of \cite{Gaberdiel:2011aa}, we can actually identify the corresponding
coset states very explicitly. Indeed, following a similar analysis for the  ${\cal N}=2$
superconformal case \cite{GK} (see also \cite{Fredenhagen:2012bw} for earlier work in this direction), 
we claim that the twisted sector ground states correspond to the 
coset representations
\be\label{twground}
\Bigl( \Lambda_+^{(m)}; \Lambda_-^{(m)}, \hat{u}^{(m)} \Bigr) \ , 
\ee
where $m$ takes the values $m=1,\ldots,N+1$, and 
\be
\Lambda_+^{(m)} = [\Lambda_1, \ldots, \Lambda_{N+1}] \ , \qquad \hbox{with} \ \Lambda_m = 0 
\ee
is a weight of $\mathfrak{su}(N+2)$ such that 
\be
\sum_{i=1}^{m-1} \Lambda_i \leq \frac{k}{2} \ , \qquad \hbox{and} \qquad
\sum_{i=m+1}^{N+1} \Lambda_i \leq \frac{k}{2} \ .
\ee
(Note that the weight $\Lambda_+^{(m)}$ is allowed at level $k$.)
Furthermore, we take $\Lambda_-^{(m)}$ to be the weight of $\mathfrak{su}(N)$ defined by
\be
\Lambda_-^{(m)} = [\Lambda_1, \ldots, \Lambda_{m-2},\Lambda_{m-1} + \Lambda_m + 
\Lambda_{m+1}, \Lambda_{m+2}, \ldots, 
\Lambda_{N+1}] \ ,
\ee
and set the $\mathfrak{u}(1)$ charge to
\be
\hat{u}^{(m)} = 2 \Bigl( \sum_{i=1}^{m-1} i \Lambda_i - \sum_{j=m+1}^{N+1} (N+2-j) \Lambda_j \Bigr) -(N + 2 - 2m) \, \Lambda_m \ . 
\ee
One easily confirms that the triplet $(\Lambda_+^{(m)}; \Lambda_-^{(m)} , \hat{u}^{(m)})$ satisfies then
the  selection rule (\ref{N4sel}). We now claim that, in the limit $k\rightarrow\infty$,
the twist corresponding to (\ref{twground}) is precisely  (for $\Lambda_m=0$)\footnote{
If $\Lambda_m\neq 0$ then in order for the representation to have finite conformal dimension in 
the $k\rightarrow \infty$ limit, $\Lambda_m \sim \sqrt{k}$, while $\Lambda_i \sim k$ for $i\neq m$. 
This general case corresponds to a twisted sector where in addition some momentum 
(proportional to $\Lambda_m$) along the
${\rm S}^3$ whose radius goes to infinity has been switched on. As mentioned earlier, we will 
not consider states with $\Lambda_m\neq 0$ in the following.} 
\be\label{twist}
\alpha = \frac{1}{N+k+2}\, \Bigl[  \sum_{i=1}^{m-1} \Lambda_i \; , \  \sum_{i=2}^{m-1} \Lambda_i \ , \ldots \ , 
\Lambda_{m-1} , \ - \Lambda_{m+1} , - \sum_{i=m+1}^{m+2} \Lambda_i \ ,  \ldots  , \ 
- \sum_{i=m+1}^{N+1} \Lambda_i \Bigr] \ . 
\ee
Note that the twist is only non-trivial if the Dynkin labels $\Lambda_i$ scale with $k$, as is also familiar
from the bosonic analysis of \cite{Gaberdiel:2011aa}.

\subsection{Comparing the Conformal Dimension}

There are various pieces of evidence in support of this claim. First of all, we can determine the conformal
dimension of the representation (\ref{twground}). The key step in this calculation is the observation that
the difference of Casimirs takes the form
\begin{eqnarray}
C^{(N+2)}\bigl(\Lambda_+^{(m)}\bigr) - C^{(N)}\bigl(\Lambda_-^{(m)}\bigr) 
& = & \frac{1}{N(N+2)} \Biggl(\,
\sum_{i=1}^{m-1} \, i \Lambda_i - \sum_{j=m+1}^{N+1} \, (N+2-j) \Lambda_j \Biggr)^2 \nonumber \\
& & 
+ \sum_{i=1}^{m-1} \, i \Lambda_i  +  \sum_{j=m+1}^{N+1} \, (N+2-j) \Lambda_j \ . 
\end{eqnarray}
In the limit $k\rightarrow\infty$, we therefore conclude from (\ref{N4hdef}) that the conformal
dimension is indeed just (\ref{ground}), with the twist $\alpha$ being given by (\ref{twist}). Note
that, for these representations, $n=0$ in (\ref{N4hdef}) since $\Lambda_-^{(m)}$ appears
in the branching of $\Lambda_+^{(m)}$ from $\mathfrak{su}(N+2)$ to $\mathfrak{su}(N)$,
as follows from the analysis of Appendix~B of \cite{GK}.

\subsection{The Fermionic Excitation Spectrum}

A somewhat more refined test is provided by calculating the fermionic excitation spectrum of these
ground states. Recall from (\ref{ffer}) that we can identify the fermionic fields with the coset
primaries $(0;{\rm f},(N+2))$ as well as their conjugates. If we apply this coset representation (or its
conjugate) to
the twisted sector ground state we will, generically, obtain $N$ different fusion products
\be\label{fermdesc}
\Bigl(\Lambda^{(m)}_+; \Lambda_-^{(m) \epsilon l},\hat{u}^{(m)} + \epsilon (N+2) \Bigr) \ , 
\ee
where  $\epsilon=\pm$ labels whether we consider the fermions or their conjugates, and 
$\Lambda^{(m) \epsilon l}$ are those representations (with $l\in \{1,\ldots, N\}$) that appear in 
\be\label{ffus}
{\rm f} \otimes \Lambda = \bigoplus_{l=1}^{N} \Lambda^{+l} \ , \qquad 
\bar{\rm f} \otimes \Lambda = \bigoplus_{l=1}^{N} \Lambda^{- l} \ . 
\ee
Indeed, we have explicitly
\be\label{fusionf}
\Lambda^{\epsilon l}_j = \left\{
\begin{array}{ll}
\Lambda_j - \epsilon \qquad & j=l-1 \\
\Lambda_j + \epsilon \qquad & j=l \\
\Lambda_j \qquad & \hbox{otherwise.}
\end{array}
\right. 
\ee
We can therefore read off the fermionic excitation spectrum of the various fermion fields by comparing
the conformal dimensions (see \cite{GK} for a similar analysis in the ${\cal N}=2$ case)
\begin{eqnarray}\label{deltah}
\delta h^{(l)} & = &  
h\Bigl(\Lambda^{(m)}_+; \Lambda_-^{(m) \epsilon l},\hat{u}^{(m)} + \epsilon (N+2) \Bigr) \ -  \ 
h\Bigl(\Lambda^{(m)}_+; \Lambda_-^{(m)},\hat{u}^{(m)}  \Bigr) \nonumber \\
& = & \frac{1}{2} + \frac{1}{N+k+2}  \Bigl[
\, \frac{\epsilon}{N}  \Bigl(\,  \sum_{i=1}^{N-1} \, i\, \tilde\Lambda_i -   \frac{\hat{u}^{(m)}}{2} \Bigr) 
- \epsilon \sum_{j=l}^{N-1} \tilde\Lambda_j  \Bigr] \nonumber \\
& & + \frac{1}{2 (N+k+2)} \bigl(- \epsilon N + 2 l \epsilon + (\epsilon - \tfrac{7}{2}) \bigr) \ ,
\end{eqnarray}
where, to simplify notation, we have set $\Lambda^{(m)}_- = \tilde\Lambda$. 
In the limit $k\rightarrow \infty$, the last line can be ignored (since none of the terms in 
the numerator can depend on $k$), and hence we get approximately
\begin{eqnarray} \label{shift}
\delta h^{(l)} & \cong & \frac{1}{2} - \frac{\epsilon}{N+k+2} \,
\Bigl[\, \sum_{j=l}^{N-1} \tilde\Lambda_j +  \frac{1}{N}  \Bigl( \, \frac{\hat{u}^{(m)}}{2} 
- \sum_{i=1}^{N-1} \, i\, \tilde\Lambda_i \Bigr) \Bigr] \nonumber \\
& = & \frac{1}{2} - \epsilon\, \alpha_{l} \ . 
\end{eqnarray}
where we have used the explicit expression for $\hat{u}^{(m)}$ in the final step, and 
$\alpha_j$ is the $j$'th component of the vector $\alpha$  in (\ref{twist}). Thus the
different fusion channels correspond to the different twisted modes, and the result is in perfect agreement
with our identification of the twists.

\subsection{The BPS Spectrum}

Finally, in analogy with the situation for the symmetric orbifold, we may expect that a certain fermionic
descendant of these twisted sector ground states should be BPS. Indeed, we should simply apply all fermions
whose $\delta h^{(l)}$ is less than $1/2$, i.e., the fundamental fermions $(\epsilon=+)$ with $l=1,\ldots, m-1$, and
the anti-fermions $(\epsilon=-)$ with $l=m+1,\ldots,N$.  Thus the relevant BPS descendant should be 
\be\label{BPSdes}
\Bigl(\Lambda_+^{(m)}; [\Lambda_1,\ldots,\Lambda_{m-2},\Lambda_{m-1} + \Lambda_{m+1} + 2, 
\Lambda_{m+2},\ldots,\Lambda_{N+1}],\hat{u}^{(m)} +  (N+2) (2m - 2 -N) \Bigr) \ .
\ee
It is not difficult to check that this state satisfies then the selection rule with
\be
\frac{|\Lambda_+|}{N+2} - \frac{|\Lambda_-|}{N} + \frac{\hat{u}}{N(N+2)} = -1 \in \mathbb{Z} \  ,
\ee
and that its quantum numbers are 
\be
l^+=0  \ , \qquad l^- = \frac{N}{2} \ , \qquad u=0 \ .
\ee
Indeed, we need $N$ fermionic descendants to obtain the denominator representation from the numerator,
and each fermion transforms in the spin $j=\frac{1}{2}$ representation of $\mathfrak{su}(2)_-$; thus the
relevant BPS bound is
\be\label{BPSbound}
h_{\rm BPS} = \frac{1}{N+k+2} \Bigl( (k+1) \frac{N}{2} + \frac{N^2}{4} \Bigr) 
= \frac{2 N (k+1) + N^2}{4 (N+k+2)} \ ,
\ee
and one checks by an explicit (albeit slightly tedious calculation) that this indeed equals the conformal
dimension of (\ref{BPSdes}). In fact, this statement is even true at finite $N$ and $k$.

\section{The Twisted Sector of the Symmetric Orbifold}\label{sec:7}

Recall from the discussion in Section~\ref{sec:light} that the twisted sectors
of the symmetric product orbifold form a subset of the twisted sectors of the continuous
orbifold. In this section we want to identify the relevant twisted sector representations
in terms of the coset language.

\subsection{Locality}\label{sec:local}

As alluded to in Section~\ref{sec:light}, it is  in general quite difficult to determine which coset 
representations are
local with respect to an extended chiral algebra. However, for the case of interest, i.e., the
extended chiral algebra (\ref{Wcharexp}), there are some simple checks we can perform.

First of all, it is easy to see that {\em all} representations of the form $(0;\Lambda)$ 
with $\Lambda$ a finite representation (i.e., made up of finitely many boxes and anti-boxes) 
are  local with respect to the extended chiral algebra in the limit $k\rightarrow \infty$. This is 
simply a consequence of the fact that, for these representations
\be
h\bigl(0;\Lambda,  |\Lambda| (N+2)\bigr) = 
n - \frac{C^{(N)}(\Lambda)}{N+k+2} -\frac{|\Lambda|^2 (N+2)}{4 N (N+k+2)} \cong n 
\ee
in the large $k$ limit, where $n$ is the excitation number that describes the level at which
$\Lambda$ appears in the vacuum representation of the numerator. Since $n$ is, by construction,
a non-negative half-integer, the conformal dimension of all of these representations is half-integer
or integer. But since the extended chiral fields from  (\ref{Wcharexp}) map a representation
of the form $(0;\Lambda)$  to another representation of the same kind, $(0;\Lambda')$, it follows that 
locality is manifest for all of these
representations. This is obviously important since, as we saw above, see eq.~(\ref{pert0}),
all of these representations actually survive in the untwisted sector of the continuous orbifold
and hence must be allowed representations of the extended theory.
\medskip

The situation is a little more interesting for the various twisted representations of the continuous
orbifold, i.e., the representations discussed in Section~\ref{sec:twisted}. A necessary condition for 
$(0;{\tiny \yng(2)})$ to be local with respect to the ground states $(\Lambda_+;\Lambda_-,\hat{u})$
is that the fusion 
\be
\bigl(\Lambda_+;\Lambda_-,\hat{u}\bigr) \otimes (0;{\tiny \yng(2)}) 
\ee
contains a representation whose conformal dimension differs from that of $(\Lambda_+;\Lambda_-,\hat{u})$
by a half-integer (or integer). This can be calculated using the same techniques as in Section~\ref{sec:twisted};
indeed, the relevant tensor product equals simply
\be
\Lambda_- \otimes {\tiny \yng(2)} = \bigoplus_{l_1\leq l_2}  \bigl( \Lambda_-^{+l_1} \bigr)^{+l_2} \ ,
\ee
and the difference in conformal dimension becomes, in the large $k$ limit, the sum 
$\delta h^{(l_1)} + \delta h^{(l_2)}$ of the two individual shifts --- we are using here the same conventions as 
in eqs.~\eqref{fusionf} -- \eqref{shift}. Thus locality with respect to $(0;{\tiny \yng(2)})$ requires that the twisted sector
ground state satisfies
\be\label{local2}
\alpha_{l_1} + \alpha_{l_2} \in \tfrac{1}{2} \mathbb{Z} \ , 
\ee
for some choice of  $l_1,l_2\in \{1,\ldots, N\}$. Condition (\ref{local2}) is obviously satisfied for the twisted sectors
of the symmetric orbifold --- for  the twisted sector associated to a single $r$-cycle permutation, the twists are 
contained, modulo one, in the set $\{\frac{m}{r}, m=0,\ldots, r-1\}$, and similarly for products of 
$r$-cycle permutations. However, for a generic twisted sector of the continuous ${\rm U}(N)$ orbifold,
this condition will {\em not} be satisfied. Thus we see that locality does impose non-trivial constraints on the
allowed representations.

The situation is similar for the $(0;{\tiny \yng(3)})$ primary in eq.~\eqref{Wcharexp}, for which instead of 
(\ref{local2}) we obtain the condition
\be\label{local3}
\alpha_{l_1} + \alpha_{l_2} + \alpha_{l_3}  \in \tfrac{1}{2} \mathbb{Z} \ , 
\ee
for some choice of  $l_1,l_2, l_3\in \{1,\ldots, N\}$. Again, this is satisfied for the  twisted sectors
of the symmetric orbifold, but in general not for a generic twisted sector of the continuous ${\rm U}(N)$ orbifold.
Finally, for the representation associated to $(0;[2,0,0,\ldots,0,1])$ we obtain instead
\be\label{local21}
\alpha_{l_1} + \alpha_{l_2} - \alpha_{l_3}  \in \tfrac{1}{2} \mathbb{Z} \ , 
\ee
for some choice of $l_1,l_2, l_3\in \{1,\ldots, N\}$. These constraints are obviously only necessary conditions
for locality ---  unfortunately, no simple sufficient condition for locality is known. It is nevertheless reassuring
that they are satisfied for the twisted sectors of the symmetric orbifold, but not for generic twisted
sectors of the continuous orbifold.

\subsection{The $2$-cycle Twist Sector}\label{2cycle}

We can also understand how specific  twisted sectors of the symmetric orbifold fit into
our picture. Let us illustrate this with the simplest example, the 
case of the $2$-cycle twist; more complicated cases could also be similarly studied, but
we have not attempted to do so.

We start by calculating the partition function of the symmetric orbifold in this sector. Using
again the techniques of \cite{Dijkgraaf:1996xw} (see also \cite{Maldacena:1999bp}) it follows that the 
corresponding  generating function equals 
\begin{eqnarray}\label{generating2}
\sum_{k=0}^{\infty} p^k Z^{({\rm 2})} ({\rm Sym}^k(X)) & = & 
p^2 {\sum_{\Delta,\bar\Delta,\ell,\bar\ell}}' \, c(\Delta,\bar\Delta, \ell,\bar\ell)\, 
q^{\frac{\Delta}{2}} \bar{q}^{\frac{\bar{\Delta}}{2}}\, y^\ell \bar{y}^{\bar{\ell}}\,  \nonumber \\
& & \qquad \times  \prod_{\Delta,\bar\Delta,\ell,\bar\ell} 
\frac{1}{(1 - p q^{\Delta} \bar{q}^{\bar\Delta} y^{\ell} \bar{y}^{\bar{\ell}} )^{ 
c(\Delta,\bar\Delta, \ell,\bar\ell) }} \ ,
\end{eqnarray}
where the prime in the sum of the first line means that we only sum over the $4$-tuples 
$(\Delta,\bar\Delta,\ell,\bar\ell)$ for which $\Delta-\bar\Delta$ is even. For the case at hand,
the result again factorises into a sum over 
chiral and anti-chiral functions. Note that for the $\mathbb{Z}_2$ 
twisted sector, the different contributions are invariant under the centraliser of the $\mathbb{Z}_2$ in 
$S_{N+1}$, which is $S_{N-1}\times S_2$. 
We will focus on the part of the expansion of (\ref{generating2}) 
where the left and right movers are separately invariant under the $S_{N-1}$. Then the only 
remaining distinction is whether they are even or odd with respect to the $S_2\cong \mathbb{Z}_2$. Both
classes of states contribute, but the overall invariance under the centraliser implies that 
the $S_2$ even/odd states  for the left-movers are coupled to  $S_2$ even/odd states of the right movers, 
respectively --- this is precisely what the condition represented by the prime in the above sum
implements. Thus we can analyse separately the $S_2$ even/odd chiral characters, and
we find in the NS sector, repeating essentially the steps of Section~\ref{sec:Wext}
\begin{eqnarray}\label{chi+}
{\cal Z}^{(2)}_{+}(q) & = & 
q^{\frac{1}{2}} \Bigl( 
(y + y^{-1}) + q^{1/2} \, \bigl( 4 y^2 + 16 + 4 y^{-2} \bigr) \nonumber \\
& & \qquad 
{}+ q^{1} \bigl( 7 y^3 + 81 y + 81 y^{-1} + 7 y^{-3} \bigr) \nonumber \\
& & \qquad {}+ q^{3/2} (8 y^4 + 218 y^2 + 580 + 218 y^{-2} + 8 y^{-4} \bigr) + \cdots \Bigr)  \label{chi2p}
\end{eqnarray}
for the $S_2$ even states of the $2$-cycle twisted sector, while the character of the 
$S_2$ odd states in the $2$-cycle twisted sector equals
\begin{eqnarray}\label{chi-}
{\cal Z}^{(2)}_{-}(q) & = & 
q^{\frac{1}{2}} \Bigl( 
2 + q^{1/2} \, \bigl( 12 y + 12 y^{-1} \bigr) \nonumber \\
& & \qquad {}
+ q^{1} \bigl( 32 y^2 + 112+ 32 y^{-2} \bigr) \nonumber \\
& & \qquad {}+ q^{3/2} (52 y^3 + 464 y + 464 y^{-1} + 52 y^{-3} \bigr) + \cdots \Bigr)  \ . \label{chi2m}
\end{eqnarray}
In order to facilitate comparison with the coset representations, let us understand the 
behaviour of the $(N+1)$ free fermions (and bosons) of the  symmetric product orbifold under the 
subgroup $S_{N-1}\times S_2\subset S_{N+1}$; it is not difficult to see that  they transform as 
\be
N+1 \cong N \oplus 1 \cong 
\Bigl[ (1_{N-1}\otimes 1_2) \oplus (1_{N-1} \otimes 1'_2) \oplus \bigl( (N-2)_{N-1} \otimes 1_2\bigr) \bigr] 
\oplus  (1_{N-1}\otimes 1_2) \ , 
\ee
where $1_L$ denotes the singlet and $1^{\prime}_L$ the alternating singlet (which is odd under the 
odd permutations), and the index labels the relevant $S_{L}$ group. The last
singlet $ (1_{N-1}\otimes 1_2)$ is just the overall singlet of $S_{N+1}$, i.e., 
the sum of all $N+1$ fermionic modes (that remains untwisted and hence half-integer moded). 
The $(1_{N-1}\otimes 1^{\prime}_{2})$ represents the fermion which is odd under the 
$S_2\cong \mathbb{Z}_2$ and which is therefore integer-moded in this sector --- in particular, it therefore
includes a zero mode. The other fermions,  namely the $(1_{N-1}\otimes 1_2)$ and the 
$\bigl((N-2)_{N-1}\otimes 1_2\bigr)$, 
being even under this $\mathbb{Z}_2$, continue to be half-integer moded. 
\smallskip

The corresponding Wolf coset representations can be read off from the analysis of Section~\ref{sec:twisted}, 
except that, as we have just seen, the $2$-cycle twist  is somewhat degenerate in that the twisted sector has
a  fermionic zero mode --- this will always be the case if the cycle has even length. As a consequence,
there is not just a unique twisted sector ground state, but rather a whole representation of the corresponding
Clifford algebra. In fact, `the' twisted sector ground state of Section~\ref{sec:twisted}, 
\be\label{tw0}
\Lambda_+ = [\tfrac{k}{2},0,0,\ldots,0] \ , \qquad 
\Lambda_-= [\tfrac{k}{2},0,0,.\ldots, 0] \ , \qquad \hat{u}  = k  \ ,
\ee
is only one of the two states contributing to the leading term in eq.~(\ref{chi2m}). Its BPS `descendant' 
of eq.~(\ref{BPSdes})\footnote{Since all $\Lambda_j$ with $j\geq m=2$ vanish, there is only one 
fundamental fermion we should apply, and hence the $+2$ in eq.~(\ref{BPSdes}) is replaced by a $+1$.}
\be\label{tw1}
\Lambda_+ = [\tfrac{k}{2},0,0,\ldots,0] \ , \qquad 
\Lambda_-= [\tfrac{k}{2}+1,0,0,.\ldots, 0] \ , \qquad \hat{u}  = k +(N+2) \ ,
\ee
is in fact degenerate in conformal dimension since the mode
of the relevant fermion is a zero mode. The state in eq.~(\ref{tw1}) has $l_{-}=\frac{1}{2}$, and it  accounts 
precisely for the leading terms in (\ref{chi2p}). Finally, applying the relevant fermionic zero mode again 
we obtain another coset primary with $l_{-}=0$ 
\be\label{tw2}
\Lambda_+ = [\tfrac{k}{2},0,0,\ldots,0] \ , \qquad 
\Lambda_-= [\tfrac{k}{2}+2,0,0,.\ldots, 0] \ , \qquad \hat{u}  = k + 2 (N+2) \ ,
\ee
which accounts for the other leading term in (\ref{chi2m}). We should mention in passing that we have the field 
identifications
\be
\Bigl( [\tfrac{k}{2},0,\ldots,0] ; [\tfrac{k}{2},0,\ldots, 0]  , k \Bigr)  \cong 
\Bigl( [0,\ldots,0,\tfrac{k}{2}] ; [0,.\ldots, 0,\tfrac{k}{2} +2]  , - k - 2 (N+2)\Bigr)
\ee
and
\be
\Bigl( [\tfrac{k}{2},0,\ldots,0] ; [\tfrac{k}{2}+2,0,\ldots, 0]  , k + 2(N+2) \Bigr)  \cong 
\Bigl( [0,\ldots,0,\tfrac{k}{2}] ; [0,.\ldots, 0,\tfrac{k}{2} ]  , - k \Bigr) \ , 
\ee
thus showing that the twisted representation with twist $\alpha=\frac{1}{2}$ is
indeed equivalent to that with twist $\alpha=-\frac{1}{2}$, as well as demonstrating
that the two representations (\ref{tw0}) and (\ref{tw2}) are on the same footing. On 
the other hand, the field identification of the coset representation (\ref{tw1}) is simply
\be
\Bigl( [\tfrac{k}{2},0,\ldots,0] ; [\tfrac{k}{2}+1,0,.\ldots, 0]  , k +N+2 \Bigr)  \cong 
\Bigl( [0,\ldots,0,\tfrac{k}{2}] ; [0,.\ldots, 0,\tfrac{k}{2} +1]  , - k - N-2 \Bigr)  .
\ee

\subsection{Comparing Characters}

The characters of the lowest coset representations equal --- more details
can be found in the ancillary file of the {\tt arXiv} submission\footnote{We thank 
Constantin Candu for helping us check these identities.}
\begin{eqnarray}
\chi_{ [k/2,0,0,\ldots,0] ; [k/2,0,0,\ldots, 0] }(q) & = & 
q^{1/2} \Bigl( 1 + 4 (y+ y^{-1}) q^{1/2} + (7y^2 + 27 + 7y^{-2}) q + 
\cdots \Bigr) \nonumber \\[2pt]
\chi_{ [k/2,0,0,\ldots,0] ; [k/2+2,0,0,\ldots, 0] }(q) & = & 
q^{1/2} \Bigl( 1 + 4 (y+ y^{-1}) q^{1/2} + (7y^2 + 27 + 7y^{-2}) q + \cdots \Bigr) \nonumber \\[2pt]
\chi_{ [k/2,0,0,\ldots,0] ; [k/2+1,0,0,\ldots, 0] }(q) & = & 
q^{1/2} \Bigl( (y+ y^{-1}) + (2 y^2 + 8 + 2 y^{-2}) q^{1/2} \nonumber \\
& & \qquad \ \  + (2y^3+26 y+26 y^{-1}+2 y^{-3}) q^1 + \cdots \Bigr) \nonumber \ .
\end{eqnarray}
Combining these results we therefore have an expansion of the form 
\begin{eqnarray}\label{cosetexp}
{\cal Z}^{(2)}_{+}(q) & = &  
\chi_{ [k/2,0,0,\ldots,0] ; [k/2+1,0,0,.\ldots, 0] }(q)  \nonumber \\
& & +\ \chi_{ [k/2,0,0,\ldots,0] ; [k/2-1,0,0,\ldots, 0] }(q)
+ \chi_{ [k/2,0,0,\ldots,0] ; [k/2+3,0,0,\ldots, 0] }(q) \nonumber \\
& & + \ \chi_{ [k/2,0,0,\ldots,0] ; [k/2,1,0,\ldots, 0] }(q) + \chi_{ [k/2,0,0,\ldots,0] ; [k/2+1,0,\ldots, 0,1] }(q) \nonumber \\
& & + \ \chi_{ [k/2,0,0,\ldots,0] ; [k/2,1,0,\ldots, 0,1] }(q) \nonumber \\
& & + \ \chi_{ [k/2,0,0,\ldots,0] ; [k/2-2,1,0,\ldots, 0,0] }(q) + \chi_{ [k/2,0,0,\ldots,0] ; [k/2+3,0,\ldots, 0,1] }(q) \nonumber \\
& & + \ \chi_{ [k/2,0,0,\ldots,0] ; [k/2-1,0,\ldots, 0,1] }(q)  + \chi_{ [k/2,0,0,\ldots,0] ; [k/2+2,1,\ldots, 0,0] }(q)  \nonumber \\
& & + \ 2 \cdot \chi_{ [k/2,0,0,\ldots,0] ; [k/2-1,2,\ldots, 0,0] }(q) 
+ 2 \cdot \chi_{ [k/2,0,0,\ldots,0] ; [k/2+1,0,\ldots, 0,2] }(q)  \nonumber \\
& & +  \ {\cal O}(q^{2}) \ , \\
{\cal Z}^{(2)}_{-}(q) & = &  
\chi_{ [k/2,0,0,\ldots,0] ; [k/2,0,0,.\ldots, 0] }(q) +
\chi_{ [k/2,0,0,\ldots,0] ; [k/2+2,0,0,.\ldots, 0] }(q)  \nonumber \\
& & + \ \chi_{ [k/2,0,0,\ldots,0] ; [k/2,0,0,\ldots, 1] }(q) + \chi_{ [k/2,0,0,\ldots,0] ; [k/2+1,1,0,\ldots, 0] }(q)  \nonumber\\
& & + \ \chi_{ [k/2,0,\ldots,0] ; [k/2-1,1,0,\ldots, 0] }(q) + \chi_{ [k/2,0,0,\ldots,0] ; [k/2+2,0,0,\ldots, 0,1] }(q)  \nonumber  \\
& & + \ \chi_{ [k/2,0,0,\ldots,0] ; [k/2-1,1,\ldots, 0,1] }(q) + \chi_{ [k/2,0,0,\ldots,0] ; [k/2+1,1,\ldots, 0,1] }(q) \nonumber  \\
& & + \ \chi_{ [k/2,0,0,\ldots,0] ; [k/2-2,0,0,\ldots, 0] }(q)  + \chi_{ [k/2,0,0,\ldots,0] ; [k/2+4,0,0,\ldots, 0] }(q) \nonumber  \\
& & + \ 2 \cdot \chi_{ [k/2,0,0,\ldots,0] ; [k/2,0,0,\ldots, 2] }(q)  +  2 \cdot \chi_{ [k/2,0,0,\ldots,0] ; [k/2,2,0,0,\ldots, 0] }(q) 
\nonumber \\
& & + \ 2 \cdot \chi_{ [k/2,0,0,\ldots,0] ; [k/2+2,0,0,\ldots, 2] }(q) + 2 \cdot \chi_{ [k/2,0,0,\ldots,0] ; [k/2-2,2,0,0,\ldots, 0] }(q)  
\nonumber \\
& & +\ {\cal O}(q^{2}) \ .  \label{cosetexp1}
\end{eqnarray}

It is possible to understand the systematics of which coset 
representations appear in eqs.~(\ref{cosetexp}) and (\ref{cosetexp1}), 
and with which multiplicity, i.e., the analogue of (\ref{vacgen}). In order to explain
this let us start with the twisted sector ground state (\ref{tw0}). The excitations are either the 
fermion zero mode (and its descendants) described below (\ref{chi-}) and identified in the coset below 
(\ref{tw0}), or the half-integer modes which transform as 
$(1_{N-1}\otimes 1_2)\oplus \bigl((N-2)_{N-1}\otimes 1_2\bigr)$ of $S_{N-1}\times S_2$. 
Each action of the former changes the parity of the state, since the fermionic zero mode is odd under the $S_2$, 
 while the latter, being even under the $S_2$, do not modify the parity. 
 
On the other hand, as can be seen from eq.~(\ref{shift}), for 
$\ell=1$ (and $\epsilon=1$) we have a zero mode, whereas for $\ell\neq 1$ we have 
$\delta h^{(\ell)} =\frac{1}{2}$ (since $\alpha_{\ell}=0$ for $\ell\neq 1$ for the state (\ref{tw0})).  Since $\ell$ labels the 
row to which extra boxes are attached, we conclude that, from the coset viewpoint,
the integer moded excitations are those  which increase the first row ($\ell =1$), while 
the half-integer modes are associated with adding boxes in the Young tableau below the first row.  
The situation for the anti-boxes (that describe the complex conjugate fermions) is the same. 
 
Let us rephrase this in the language of Dynkin labels. All the coset representations
that are of relevance have  $\Lambda_+= [\frac{k}{2}, 0 \ldots ,0]$, while $\Lambda_-$ is of the form
$\Lambda_- = [\frac{k}{2} +l_0, \Lambda^{\prime}]$, where $l_0\in\mathbb{Z}$ and
$\Lambda^{\prime}$ denotes the remaining $(N-2)$ Dynkin labels. 
Changing $l_0$, while keeping $\Lambda^{\prime}$ fixed, corresponds to adding an integer moded
fermion (that is odd under the $S_2\cong\mathbb{Z}_2$). When we 
add a box below the first row, we not only modify the first few Dynkin labels of $\Lambda^\prime$,  
we automatically also shift $l_0$ by one; on the other hand, if we add an anti-box, we only
modify the last few Dynkin labels of $\Lambda^\prime$,  but this does not influence 
$l_0$.\footnote{Note that since the underlying representations are really ${\rm U}(N)$ representations, 
an anti-box does not coincide with the ${\rm SU}(N)$ Dynkin label $[0,\ldots,0,1]$. However,
since we have not kept track of the ${\rm U}(1)$ charge in our notation and since we are working with
small excitations, we have used the convention that the first few Dynkin labels refer to boxes, while
the last few Dynkin labels correspond to anti-boxes. We hope this will not create undue confusion.} Thus we conclude that 
the parity with respect to $S_2\cong \mathbb{Z}_2$ of a given state (relative to the ground state
state (\ref{tw0})) is 
\be
P= l_0+\sum_{i} \Lambda^{\prime}_i \qquad \hbox{mod $2$} \ ,
\ee
where the sum runs only over the `first few' Dynkin labels that correspond to the addition of 
boxes. In particular, $P$ is insensitive to the number of anti-boxes (see footnote 12). 
This gives the selection rule for which states appear in (\ref{chi+}) and (\ref{chi-}), respectively: the ones 
with $P=0\  ({\rm mod}\ 2)$ appear in (\ref{chi-}), while those that satisfy  
$P=1 \ ({\rm mod}\  2)$ appear in (\ref{chi+}). An inspection of (\ref{cosetexp}) and (\ref{cosetexp1})
bears this out. 

In order to understand the multiplicities with which the 
representations actually appear, we now recall that (\ref{cosetexp}) and (\ref{cosetexp1}) count the states that 
are $S_{N-1}$ invariant (separately for left- and right-movers). Since 
the modes corresponding to the first row are singlets with respect to $S_{N-1}$, only
the Dynkin labels in $\Lambda^\prime$ matter. They describe
a representation of ${\rm SU}(N-1)$, and we therefore need to determine the branching
rules of $\Lambda^\prime$ with respect to the embedding 
$S_{N-1}\subset {\rm U}(N-2)\subset {\rm SU}(N-1)$. Under this embedding, both the fundamental
and anti-fundamental representation of ${\rm SU}(N-1)$ decompose as 
\be\label{decomp}
({\bf N-1}) \ \rightarrow \ (N-2)_{N-1} \oplus 1_{N-1}  \ , \qquad 
\overline{({\bf N-1})}  \ \rightarrow \ (N-2)_{N-1} \oplus 1_{N-1}  \ .
\ee
Thus relative to the analysis of Section~\ref{sec:Wext} we now get additional singlets coming from
the explicit singlet representation appearing in (\ref{decomp}). In particular, we get a singlet from
adding a single box (or anti-box), see the contribution in the third line of (\ref{cosetexp}) and
the second and third line of (\ref{cosetexp1}). The symmetric product of two boxes (or anti-boxes) 
contains the singlet with multiplicity $2$ --- see the last line of (\ref{cosetexp})  and the last two
lines of (\ref{cosetexp1}) --- while the anti-symmetric product does not contain any singlet, etc. 
More generally, the decomposition of this twisted sector thus takes the form
\be\label{twistgen}
{\cal Z}^{(2)}_{\pm}(q,y) 
= \sum_{\Lambda^{\prime}, l_0}  \,\delta^{(2)}_{\pm}( P) \,  \,
\widetilde{n} (\Lambda^{\prime}) \, 
\chi_{([\frac{k}{2}, 0 \ldots ,0];[\frac{k}{2}+l_0, \Lambda^{\prime}])}(q,y) \ ,
\ee 
where $\widetilde{n}(\Lambda^{\prime})$ is the multiplicity of singlets of $S_{N-1}$ in 
$\Lambda^{\prime}$, and the sum over $l_0$ is restricted by the parity requirement that
 $P=0\  ({\rm mod}\ 2)$ for ${\cal Z}_-$, and  $P=1\  ({\rm mod}\ 2)$ for ${\cal Z}_+$ --- this is what
 is imposed by the factor $\delta^{(2)}_{\pm}( P)$.

\section{Concluding Remarks}\label{sec:8}

We close with some comments, elaborating on the meaning of some of our results as well as 
raising questions that we feel could be addressed in the near future. 

The main thrust of our paper has been to understand the exact relation between the higher spin theory 
(and its symmetries) on ${\rm AdS}_3$, and string theory as captured by the symmetric product CFT. We found
in particular
\begin{itemize}
\item
that the (perturbative) Vasiliev theory is a subsector of the full string theory in a precise sense --- it is the 
untwisted sector (\ref{pert0}) of the continuous orbifold $({\mathbb T}^4)^{N+1}/{\rm U}(N)$ which is a 
subsector of the untwisted sector of $\bigl( \mathbb{T}^4 \bigr)^{N+1}/ S_{N+1}$. 
\item
that the full partition function (not just some index) of the symmetric product CFT can be 
written as a non-diagonal modular invariant of the ${\cal W}_{\infty}[0]$ algebra, the
chiral algebra of the continuous orbifold. This implies, in particular, that we can assemble the full spectrum 
of string theory (at the tensionless point) in terms of representations of the super 
${\cal W}_{\infty}[0]$ algebra. 
\item
that the symmetry algebra of the string theory is a huge extension of the super ${\cal W}_{\infty}$ algebra 
by an infinite number of nontrivial primaries (\ref{vacgen})  that can be characterised in terms
of the branching rules under $S_{N+1}\subset {\rm U}(N)$. This reflects, in a very precise manner, 
that string theory at the tensionless
point contains many more massless higher spin fields than those that are captured by the Vasiliev theory (which
only has one massless higher spin ${\cal N}=4$ multiplet for each integer spin).
\end{itemize} 
Let us frame these findings in terms of the general expectations from the  AdS/CFT 
correspondence in the tensionless limit.  

The symmetric product CFT is associated with the gauge theory of the D1-D5 system and thus consists of 
adjoint (and bi-fundamental) valued fields. The free  fermions and bosons of the symmetric product description 
can then be interpreted as the diagonal or Cartan elements of these adjoint fields.
Coset CFTs, on the other hand, are associated with fields in the fundamental representation, 
and the Vasiliev theory is supposed to describe the gauge fields dual to bilinears of these fields. 
While the Cartan elements are not directly related to the basis vectors of the fundamental representation, 
they are essentially the same in number (except for a shift by one which is the reason why the $(N+1)$'st
symmetric group appears for a ${\rm U}(N)$ vector model, see Appendix~\ref{app:embedding}). 
Thus, the `gauge-invariant' singlet 
states of the vector model can be put in correspondence with certain bilinears of the adjoint theory 
\be\label{fundadj}
\sum_i \psi^\ast_i\, D \,\psi_i \ \longleftrightarrow\  {\rm Tr}(\Psi\, D\, \Psi) \ .
\ee
Here, the fields on the left-hand-side transform in the fundamental or anti-fun\-da\-men\-tal representation
(with $i$ being a vector index), while on the right-hand-side, $\Psi$ is adjoint-valued; 
furthermore $D$ stands for any differential operator.  From this point of view it is then natural
that the Weyl group $S_{N+1}$ acts as a subgroup of ${\rm U}(N)$; in particular, this ${\rm U}(N)$
should not be identified directly with the gauge group of the D1-D5 system. 

In a free gauge theory,  bilinear currents as in the RHS of (\ref{fundadj}) form a closed subsector (under the OPE) of the full set of single trace 
operators \cite{Mikhailov:2002bp}. This is exactly what we found above: the untwisted sector of the continuous 
orbifold is closed under the OPE and thus a consistent subsector.\footnote{Here by consistency we mean consistency
on the sphere --- obviously this subsector is not consistent on the torus since the corresponding
partition function is not modular invariant by itself.}
\smallskip

In two dimensions, it is possible to have many more conserved currents in an adjoint theory (in the free limit) 
than the above bilinears. This is because in $d=2$ the condition on the dimension and spin for a conserved 
current, $\Delta =s$, does not constrain the currents to be bilinears in the fields unlike in higher 
dimensions.\footnote{We thank Shiraz Minwalla for discussions on this point.}
This is also what we see in the symmetric product. We have a large chiral algebra which organises itself 
into representations of the higher spin algebra (or rather its ${\cal W}_{\infty}$ extension). 
Using the explicit form of the additional primaries in Section~\ref{4.2.}, we can also see which of these are single 
particle and which are multi particle. For instance, we could consider the current primaries  in (\ref{newp4}) 
and view them schematically, using the identification with Cartan elements, as contributions of the form 
${\rm Tr}(\Psi^{\alpha}\Psi^{\beta}\partial\Phi^{a}\partial\Phi^{b})$  and  
${\rm Tr}(\Psi^{\alpha}\Psi^{\beta}){\rm Tr}(\partial\Phi^{a}\partial\Phi^{b})$ respectively. We thus have one 
single trace operator, and one double trace operator at this order. More generally, in the expansion (\ref{vacgen}), 
amongst the multiplicity of $S_{N+1}$ singlets we have contributions that come from products of smaller $S_{N+1}$ 
singlets, and those which are not decomposable in this manner (see Appendix~\ref{app:singlets}); it is only the latter that correspond to the single trace operators. 

Given all the additional currents in (\ref{vacgen}) together with the above identification of single and 
multitrace operators, we can, in principle, write down the generators of the stringy symmetry
algebra, i.e., the single trace primaries that generate the full algebra upon taking products. Their
OPEs can in principle be calculated using the free fermion and boson picture, and it would be very
interesting if one could characterise the resulting structure in a useful manner. For example, one
may hope to generalise the analysis of \cite{Beccaria:2014jra}  and enumerate the 
parameters that characterise this huge extension of ${\cal W}_{\infty}[0]$. It would also be very interesting to 
estimate, for large conformal dimensions, the number of single particle fields one has to add at each 
conformal weight. From the general correspondence with adjoint valued fields one might expect a 
Hagedorn growth. 

This stringy algebra also governs how the nontrivial primaries of the symmetric orbifold (in both untwisted and
twisted sectors) are organised in terms of coset representations. As explained in Section~\ref{5.1},
locality with respect to the extended symmetry algebra rules out the coset representations which 
correspond to the light states present in the diagonal modular invariant. This is in accord with the 
fact that light states do not arise in a gauge theory with adjoint fields whereas they are ubiquitous 
in a theory with fundamental fields on a space which allows nontrivial gauge holonomies 
\cite{Banerjee:2012gh, Banerjee:2013nca}. 

More interesting from the point of view of string theory is the decomposition of representations such 
as the ones studied in Sections~\ref{4.3} and \ref{2cycle}. We should view each of the expansions in 
(\ref{fundgen}) and (\ref{twistgen}) as a single character of the stringy ${\cal W}$-algebra, written in terms of 
${\cal W}_{\infty}[0]$ representations. The particular cases studied here contain the marginal deformations 
of the symmetric product. With the results of Sections~\ref{4.3} and \ref{2cycle}, we can now see the stringy 
multiplet that they are part of.  
The organisation of stringy states into multiplets of the higher spin symmetry algebra was something 
that was proposed for the free 4d ${\cal N}=4$ Yang-Mills spectrum in 
\cite{Bianchi:2003wx, Beisert:2003te, Beisert:2004di}, and there are some interesting similarities with what 
we find --- in particular, the representations are also organised in terms of Young tableaux of the symmetric 
group \cite{Beisert:2004di}. We should note, however, that in our setup we can actually organise
the states in representations of the extended stringy algebra (written in terms of the 
asymptotic higher spin algebra ${\cal W}_\infty[0]$), not just the original Vasiliev higher spin algebra
(which would just be ${\rm shs}_2[0]$ in our case). Nevertheless, the close similarities are worth exploring further as also any potential relation to the multi particle higher spin algebras proposed in \cite{Vasiliev:2012tv}.

The fact that the marginal operators are in non-trivial representations of the higher spin symmetry algebra 
has an important consequence. It implies that deforming the symmetric product CFT by 
these operators corresponds to giving a vev to bulk fields charged under the higher spin symmetry algebra. 
For the marginal operators from the twisted sector we should expect that they will give rise to 
a higg\-sing of the higher spin symmetry (and thus the stringy symmetry as well), i.e., that they describe
deformations that go away from the tensionless point; it would be very
interesting to check this, using the techniques of \cite{Gaberdiel:2013jpa}. 
 This is in the spirit of the general belief about the unbroken symmetric phase of string theory --- 
 see, for instance, \cite{Gross:1988ue, Witten:1988zd, Moore:1993qe, Sagnotti:2011qp}, and more recently
 in the context of AdS  \cite{Maldacena:2012sf, Vasiliev:2012tv}. 
We see that here, as is also expected in the ${\cal N}=4$ theory, the breaking is a classical effect 
from single trace operators.
In the context of the ABJ 
embedding of higher spin theory on ${\rm AdS}_4$ \cite{Chang:2012kt} one could also view this breaking as coming from the boundary 
conditions on the bulk fields  --- this arises as a one loop effect in the bulk 
Vasiliev theory. This is related to the picture of 
\cite{Chang:2012kt}, 
suggesting that the string states are being built up from non-abelian Vasiliev bits, which does not seem
to have any immediate analogue in the present case.
It would be very important to understand the similarities and 
differences with these higher dimensional cases.\footnote{It is also intriguing that extended chiral 
${\cal W}$-algebras have recently shown up in the description of supersymmetric sectors of ${\cal N}=2$ 
and ${\cal N}=4$ supersymmetric gauge theories in 4d \cite{Beem:2013sza}
(as well as in 6d \cite{Beem:2014kka}). Though these algebras have negative central charge etc., 
it might be worth understanding whether there is any precise relation between the 4d and the 2d cases.}

One of the potential payoffs from the identification of the stringy symmetries at the tensionless point is 
gaining a quantitative understanding of the broken symmetry phase. In our particular case it does not 
seem to be completely unrealistic that this could be realised. In the most optimistic scenario, one may be 
able to describe the spectrum and correlation functions of the D1-D5 system away from the symmetric 
product point without the need of having to concentrate on BPS protected quantities. Any relationship here 
to the  integrability of the string worldsheet theory, see e.g., \cite{OhlssonSax:2011ms,Sax:2012jv}
and \cite{Sfondrini:2014via} for a recent review, may also be useful for this.

One might also take encouragement from the results here to look for a similar stringy reorganisation of the 
chiral algebra and the spectrum of the `strange metal' CFTs 
\cite{Gopakumar:2012gd, Isachenkov:2014zua}. These are ${\cal N}=2$ theories and  one of the 
interesting examples amongst the general class of stringy cosets, see also 
\cite{Beccaria:2013wqa,Ahn:2013ota,Creutzig:2013tja, Candu:2013fta, Creutzig:2014ula} for 
other cases and recent discussions. 
Perhaps this will also help in identifying the dual string backgrounds. 

Another interesting direction concerns the case of ${\rm AdS}_3\times {\rm S}^3\times {\rm K3}$, 
for which one may try to identify a suitable higher spin theory and relate it to the symmetric orbifold on
K3, see \cite{BGP}. 

Finally, to return to one of the motivations of the present investigation, it is natural to believe that we may be
able to construct the CFT dual to string theory on  
${\rm AdS}_3\times {\rm S}^3\times {\rm S}^3\times {\rm S}^1$ (see \cite{Tong:2014yna} for a recent proposal)
using similar ideas.
To do that we need to go away from the $k\rightarrow \infty$ limit. While it is not immediately
obvious how to find the correct non-diagonal modular invariant that should describe the stringy
spectrum, this is at least a rather novel viewpoint for approaching this problem (and one that may
ultimately lead to success).

\section*{Acknowledgements}

We thank Ofer Aharony, Marco Baggio, 
Micha Berkooz, Constantin Candu, Justin David, Peter Goddard, Maximilian Kelm, Finn Larsen,
Juan Maldacena, Alex Maloney, Gautam Mandal, Shiraz Minwalla, 
Cheng Peng, Sanjaye Ramgoolam, and Ashoke Sen for useful conversations.
We are indebted to the Abdus Salam ICTP in Trieste for providing us with an opportunity to meet in person 
and a stimulating environment during a crucial phase of this work; we also thank the Weizmann Institute, 
Rehovot for hospitality. MRG acknowledges in addition the hospitality of 
Chulalongkorn University, Bangkok, the Solvay Institute, Brussels,  
Warsaw University, as well as MIT. RG acknowledges the hospitality of the 
International Centre for Theoretical Sciences (ICTS)-TIFR, Bangalore during the completion of this work. 
He is also unstintingly supported by the people of India through their commitment to basic scientific research.

\appendix

\section{The Large ${\cal N}=4$ Algebra and its Contraction}\label{app:susy}

The commutation and anti-commutation relations of the 
large ${\cal N}=4$ superconformal algebra $A_{\gamma}$ are
\cite{Sevrin:1988ew,Schoutens:1988ig,Spindel:1988sr,Goddard:1988wv,VanProeyen:1989me,Sevrin:1989ce}
\begin{eqnarray}
{}[U_m,U_n] & = &  \tfrac{k^+ + k^-}{2} \, m \, \delta_{m,-n}  \label{A1} \\[2pt]
{}[A^{\pm, i}_m, Q^a_r] & = & {\rm i}\, \alpha^{\pm\, i}_{ab} \, Q^b_{m+r}   \label{A2} \\[4pt]
{} \{Q^a_r,Q^b_s \} & = &  \tfrac{k^+ + k^-}{2} \, \, \delta^{ab} \, \delta_{r,-s}  \label{A3} \\[2pt]
{}[A^{\pm, i}_{m}, A^{\pm, j}_{n} ] & = &  \tfrac{k^\pm}{2}\, m \, \delta^{ij}\, \delta_{m,-n} 
+ {\rm i}\, \epsilon^{ijl}\, A^{\pm, l}_{m+n}  \label{A4} \\[2pt]
{} [U_m,G^a_r] & = & m \, Q^a_{m+r}  \label{A5} \\[2pt]
{}[A^{\pm, i}_{m},G^a_r] & = &  {\rm i} \,\alpha^{\pm\, i}_{ab} \,G^b_{m+r}  
\mp \tfrac{2 k^\pm }{k^++k^-}\, m \, \alpha^{\pm\, i}_{ab}\, Q^b_{m+r}  \label{A6} \\[2pt]
{} \{Q^a_r,G^b_s\} & = & 2\,  \alpha^{+\, i}_{ab}\, A^{+, i}_{r+s} - 2\,  \alpha^{-\, i}_{ab}\, A^{-, i}_{r+s} + \delta^{ab} \,
U_{r+s}  \label{A7} \\[4pt]
{} \{G^a_r,G^b_s\} & = & \tfrac{c}{3}\, \delta^{ab}\, (r^2 - \tfrac{1}{4}) \delta_{r,-s} 
+ 2\, \delta^{ab}\, L_{r+s} \nonumber \\
& & \ + 4\, (r-s)\, \left(\gamma \,  {\rm i}\, \alpha^{+\, i}_{ab}\, A^{+, i}_{r+s} 
+ (1-\gamma) \, {\rm i}\, \alpha^{-\, i}_{ab}\, A^{-, i}_{r+s} \right) \ ,
 \label{A8}
\end{eqnarray}
where the levels of the two $\mathfrak{su}(2)$ algebras are $k^+$ and $k^-$, and we define the 
$\gamma$ parameter by
\be
\gamma = \frac{k^-}{k^++k^-} \ . 
\ee
In the coset realisation, the levels take the values $k^+ = k+1$ and $k^- = N+1$, respectively, 
and hence $\gamma$ equals the 't~Hooft parameter $\lambda$
\be
\gamma = \lambda = \frac{N+1}{N+k+2} \ .
\ee
In the limit $k^+ = k+1\rightarrow \infty$ the large ${\cal N}=4$ superconformal algebra (\ref{A1}) -- (\ref{A8}) 
contracts to the small
superconformal ${\cal N}=4$ algebra together with $4$ free bosons and fermions. Indeed,
as already explained in \cite{Gukov:2004fh}, we need to rescale the generators whose central term 
is proportional to $k^+$; this requires that we define
\be
\hat{Q}^a_r = \frac{1}{\sqrt{k^+ + k^-}} \, Q^a_r \ , \quad
\hat{U}_m =  \frac{1}{\sqrt{k^+ + k^-}} \,  U_m \ , \quad
\hat{A}^{+,i}_m = \frac{1}{\sqrt{k^+}} \, A^{+,i}_{m} \ \ \ (\hbox{for $m\neq 0$}) \ .
\ee
Rewriting the algebra in terms of $\hat{Q}^a_r$, $\hat{U}_n$ and $\hat{A}^{+,i}_{n}$, we find
that the algebra contains in the limit $k^+\rightarrow \infty$ the subalgebra generated by 
\be
L_n \ , \quad G^a_r \ , \quad A^{-,i}_{n} \ , 
\ee
with (anti-)commutation relations
\begin{eqnarray}
{}[A^{-, i}_{m}, A^{-, j}_{n} ] & = &  \tfrac{k^-}{2}\, m \, \delta^{ij}\, \delta_{m,-n} 
+ {\rm i}\, \epsilon^{ijl}\, A^{-, l}_{m+n}  \\[2pt]
{}[A^{-, i}_{m},G^a_r] & = &  {\rm i} \,\alpha^{-\, i}_{ab} \,G^b_{m+r}   \\
{} \{G^a_r,G^b_s\} & = &  \tfrac{c}{3}\, \delta^{ab}\, (r^2 - \tfrac{1}{4}) \delta_{r,-s} 
+ 2\, \delta^{ab}\, L_{r+s} + 4\, (r-s)\, {\rm i}\, \alpha^{-\, i}_{ab}\, A^{-, i}_{r+s} \ ,
\end{eqnarray}
as well as the usual commutation relations with the Virasoro generators $L_n$. These modes therefore define
the small ${\cal N}=4$ algebra. The additional modes 
\be
\hat{Q}^a_r \ , \qquad \Bigl[ \hat{U}_n \ , \ \ \hat{A}^{+,i}_{n}  \  (i=1,2,3) \Bigr] \ \ (n\neq 0)
\ee
form the non-zero modes of  $4$ free fermions and $4$ free bosons, respectively. Finally
the zero modes $A^{+,i}_0$ form a global custodial $\mathfrak{su}(2)$ symmetry.

\section{Coset Representations and Characters}

The cosets furnish representations of the large ${\cal N}=4$ superconformal algebra extended by the 
${\cal W}$-symmetry currents. As mentioned in Section~\ref{sec:2}, they are labelled by 
$(\Lambda_+;\Lambda_-,\hat{u})$, and subject to the selection rule 
\be\label{N4sel}
\frac{|\Lambda_+|}{N+2} - \frac{|\Lambda_-|}{N} + \frac{\hat{u}}{N(N+2)} \in \mathbb{Z} \ . 
\ee
The field identification takes the form
\be
(\Lambda_+;\Lambda_-,\hat{u}) \cong \Bigl(J^{(N+2)}\Lambda_+; J^{(N)}\Lambda_-,\hat{u}+2(N+k+2) \Bigr) \ , 
\ee
where $J^{(L)}$ denotes the usual outer automorphism of $\mathfrak{su}(L)$, i.e., it maps 
\be
\Lambda = [\Lambda_0;\Lambda_1,\ldots,\Lambda_{L-1}] \ \mapsto  \ 
J^{(L)}\, \Lambda = [\Lambda_{L-1}; \Lambda_0,\Lambda_1,\ldots,\Lambda_{L-2}] \ . 
\ee
Note that this automorphism has order $N(N+2)$, since $\kappa=2N(N+2)(N+k+2)$.

\subsection{Character Formulae}\label{app:coset}

The coset character of the representation $(0;\Lambda)$  can be written, 
for $k\rightarrow \infty$ (and sufficiently large $N$) as 
\be
\chi_{(0;\Lambda)}(q,y) = \chi^{({\rm wedge})}_{(0;\Lambda)}(q,y) \,\cdot  \, \chi_0(q,y) \ , 
\ee
where $\chi_0(q,y)$ is the vacuum character of the coset ${\cal W}$-algebra (including the free fermions)
\be
\chi_0(q,y) = \prod_{n=1}^{\infty} (1 + y q^{n-1/2})^2 \, (1 + y^{-1} q^{n-1/2})^2 
 \prod_{s=1}^{\infty} 
\prod_{n=s}^{\infty} \frac{ (1 + y q^{n+1/2})^4  (1 + y^{-1} 
q^{n+1/2})^4 }{ (1 - q^n)^6 (1 - y^2 q^n) (1 - y^{-2} q^n)}  \ .
\ee 
The first few wedge characters equal explicitly\footnote{We thank Constantin Candu for providing us
with a Mathematica notebook to calculate these characters.}
\begin{eqnarray}
\chi^{({\rm wedge})}_{(0;0)}(q,y) & = & 1 \ , \nonumber \\
\chi^{({\rm wedge})}_{(0;[1,0,..0])}(q,y) & = & \frac{q^{1/2}}{(1-q)}\,  \bigl(y+y^{-1}  +  2 q^{1/2}\bigr)\ ,  \label{1000} \\
\chi^{({\rm wedge})}_{(0;[2,0,..0])}(q,y) & = & \frac{q}{(1-q)(1-q^2)}\, (1 + y q^{1/2})^2 \, (1+y^{-1} q^{1/2})^2   \ , 
\label{2000} \\
\chi^{({\rm wedge})}_{(0;[0,1,0,..0])}(q,y) & = & \frac{q}{(1-q)(1-q^2)}\,  \bigl(
(y^2+1+y^{-2}) + 2 q^{1/2} (y+y^{-1}) \nonumber \\
& & \quad  + 2 q^1 + 2 q^{3/2} (y+y^{-1}) + 3q^2 \bigr) \ , \nonumber  \\[4pt]
\chi^{({\rm wedge})}_{(0;[3,0,..0])}(q,y) & = &  \frac{q^2}{(1-q)(1-q^2)(1-q^3)}\, 
\Bigl( 2 + 4 (y+y^{-1}) q^{1/2} \nonumber \\ 
& &  \quad + (8+2 (y^2+y^{-2})) q + 5(y+y^{-1}) q^{3/2}   + (6+2 (y^2+y^{-2})) q^2  \nonumber \\
& & \quad + (y^3 + 5 y + 5 y^{-1}+y^{-3}) q^{5/2} \nonumber \\
& & \quad + (4 + 2 y^2+ 2 y^{-2}) q^3  + (y+y^{-1}) q^{7/2} \Bigr) \ , \label{3000} 
\end{eqnarray}
\begin{eqnarray}
\chi^{({\rm wedge})}_{(0;[0,0,1,0,..0])}(q,y) & = &  \frac{q^{3/2}}{(1-q)(1-q^2)(1-q^3)}\, 
\Bigl( (y^3+y+y^{-1}+y^{-3}) \nonumber \\
& & \quad + (2+2 y^2+2 y^{-2}) q^{1/2}  + (2 y+2 y^{-1})q  + (4 + 2 y^2 + 2 y^{-2}) q^{3/2}  \nonumber \\
& & \quad + (5 y+5 y^{-1}) q^2  + (6 + 2 y^2 + 2 y^{-2}) q^{5/2}  \nonumber \\
& & \quad + (4 y + 4 y^{-1}) q^3 + 4 q^{7/2} + (3 y+3 y^{-1}) q^4 + 4 q^{9/2}  \Bigr)  \ , 
\nonumber   \\
\chi^{({\rm wedge})}_{(0;[1,1,0,..0])}(q,y) & = &  \frac{q^{3/2}}{(1-q)^2 (1-q^3)}\, 
\Bigl( (y+y^{-1}) + (4 +2 y^2 + 2 y^{-2}) q^{1/2}  \nonumber \\
& &      + (y^3 + 5 y + 5 y^{-1} + y^{-3}) q   + (6 +2 y^2 + 2 y^{-2}) q^{3/2}  \nonumber \\
& & + (5 y + 5 y^{-1}) q^2  + (8 + 2 y^2 + 2 y^{-2}) q^{5/2} \nonumber \\
& & + (4 y + 4y^{-1}) q^3 + 2 q^{7/2}  \Bigr) \ .  \nonumber 
\end{eqnarray}
More explicit expressions can be found in the ancillary file of the {\tt arXiv} submission.

We also need the wedge character of the representations that involve boxes as well as anti-boxes,
e.g., 
\be
\chi^{({\rm wedge})}_{(0;[2,0,..1])}(q,y)  = \chi^{({\rm wedge})}_{(0;[2,0,..0])}(q,y)  \cdot 
\chi^{({\rm wedge})}_{(0;[0,0,..0,1])}(q,y)  \ , 
\ee
as well as its complex conjugate. Note that the wedge characters are charge-con\-ju\-ga\-tion invariant, i.e., 
\be
\chi^{({\rm wedge})}_{(0;[0,0,..0,1])}(q,y) = \chi^{({\rm wedge})}_{(0;[1,0,..0,0])}(q,y) \ . 
\ee

\subsection{A Non-Trivial Extended Representation from the Untwisted Sector}\label{app:untwist}

In this section we give an explicit formula for the first non-trivial character that appears in the 
untwisted sector, as well as its expression in terms of the continuous orbifold characters. 
In order to work out the expression for  (\ref{chi1}) we recall that the chiral NS sector 
partition function of $\mathbb{T}^4$ equals (again ignoring the $q^{-1/4}$ prefactor that 
comes from the central charge)
\begin{eqnarray} \label{E1}
Z_{\rm chiral}^{({\rm NS})} (\mathbb{T}^4) (q,y) & =  & 
1 + (2 y + 2y^{-1}) q^{1/2} + (y^2 + 8 + y^{-2}) q^1 \nonumber\\
& & + \ (12 y + 12 y^{-1}) q^{3/2} + (8 y^2 + 39 +  8 y^{-2}) q^2 \\
& & + \ (2 y^3 + 56 y + 56 y^{-1} + 2 y^{-3}) q^{5/2} 
+ (39 y^2 + 152 +  39 y^{-2}) q^3  \nonumber \\
& &  + \ {\cal O}(q^{7/2}) \ .  \nonumber
\end{eqnarray}
Then it follows from (\ref{chi1}) that 
\begin{eqnarray}\label{E2}
{\cal Z}_1(q,y) & = &  (2 y + 2y^{-1}) q^{1/2} + (5 y^2 + 16 + 5 y^{-2}) q^1 \nonumber \\
& & + \ (6 y^3 + 58 y + 58 y^{-1} + 6 y^{-3}) q^{3/2} \nonumber \\
& & + \  (6 y^4 + 128 y^2 + 315 +  128 y^{-2} + 6y^{-4}) q^2 \\
& & + \ (6 y^5 +198 y^3 + 1030 y + 1030 y^{-1} + 198 y^{-3} + 6y^{-5}) q^{5/2} \nonumber \\
& & + \ (6 y^6 + 240 y^4 + 2290 y^2 + 4724 +2290 y^{-2} +240 y^{-4} + 6 y^{-6}) q^3 \nonumber \\
& & + \ {\cal O}(q^{3}) \ . \nonumber
\end{eqnarray}
This is now to be compared with the expression in terms of continous orbifold characters,
i.e., the RHS of (\ref{fundgen}). The multiplicities $n_1(\Lambda)$ can be computed using 
similar ideas as those of Appendix~(\ref{app:singlets}); this leads to the expansion 
\begin{eqnarray}\label{Wcharexp2}
{\cal Z}_1(q,y) & = &  \chi_{(0;[1,0,...,0])}(q,y)  +  \chi_{(0;[0,...,0,1])}(q,y) + \chi_{(0;[1,0,...,0,1])}(q,y) 
\nonumber \\
& & + \ \chi_{(0;[2,0,...,0])}(q,y) + \chi_{(0;[0,0,...,0,2])}(q,y) \nonumber \\
& & + \ \chi_{(0;[1,1,0...,0])}(q,y)  +  \chi_{(0;[0,...,0,1,1])}(q,y) \nonumber \\
& & +\  2\cdot\chi_{(0;[2,0,...,0,1])}(q,y) + 2\cdot\chi_{(0;[1,0,0,...,0,2])}(q,y) \nonumber \\
& & + \ \chi_{(0;[0,2,0,...0,0])}(q,y) +  \chi_{(0;[0,0,...0,2,0])}(q,y) \nonumber  \\
& & + \ 2\cdot \chi_{(0;[3,0,...,0,0])}(q,y) + 2\cdot \chi_{(0;[0,0,0,...,0,3])}(q,y) \nonumber \\
& & + \  2\cdot \chi_{(0;[1,1,0...,0,1])}(q,y)  +  2\cdot \chi_{(0;[1,0,...,0,1,1])}(q,y) \nonumber \\
& & + \ 5\cdot \chi_{(0;[2,0,...,0,2])}(q,y) + \nonumber \\
& & + \ \chi_{(0;[0,1,0...,0,2])}(q,y)  +  \chi_{(0;[2,0,...,0,1,0])}(q,y) \nonumber \\
& & + \ 2\cdot \chi_{(0;[2,1,0,...,0])}(q,y) + 2\cdot \chi_{(0;[0,...,0,1,2])}(q,y) \nonumber \\
& & + \ \chi_{(0;[0,1,1,0,...,0])}(q,y) + \chi_{(0;[0,...,0,1,1,0])}(q,y) \nonumber \\
& & + \ 3\cdot \chi_{(0;[0,2,0,...,0,1])}(q,y) + 3\cdot \chi_{(0;[1,0,...,0,2,0])}(q,y) \nonumber \\
& & + \ 4\cdot \chi_{(0;[3,0,...,0,1])}(q,y) + 4\cdot \chi_{(0;[1,0,0,...,0,3])}(q,y) \nonumber \\
& & + \  5\cdot \chi_{(0;[1,1,0,...,0,2])}(q,y) +  5\cdot \chi_{(0;[2,0,...,0,1,1])}(q,y)  \nonumber \\
& & + \  \chi_{(0;[0,1,0...,0,1,1])}(q,y)  +   \chi_{(0;[1,1,0,...,0,1,0])}(q,y) \nonumber \\
& & +  \ 3 \cdot \chi_{(0;[4,0,...,0,0])}(q,y) + 3 \cdot \chi_{(0;[0,0,0,...,0,4])}(q,y) \nonumber \\
& & + \ 3 \cdot \chi_{(0;[1,2,0,...,0])}(q,y) + 3 \cdot \chi_{(0;[0,...,0,2,1])}(q,y) \nonumber \\
& & + \  \chi_{(0;[0,0,2,0,...,0])}(q,y) +  \chi_{(0;[0,...,0,2,0,0])}(q,y) \nonumber \\
& & + \ 4\cdot \chi_{(0;[2,1,0,...,0,1])}(q,y) + 4\cdot \chi_{(0;[1,0,...,0,1,2])}(q,y) \nonumber \\
& & + \ 2\cdot \chi_{(0;[0,1,1,0,...,0,1])}(q,y) + 2\cdot \chi_{(0;[1,...,0,1,1,0])}(q,y) \nonumber \\
& & + \  \chi_{(0;[1,0,1,0,...,0,2])}(q,y) +  \chi_{(0;[2,0,...,0,1,0,1])}(q,y)  \nonumber \\
& & + \ 7\cdot \chi_{(0;[0,2,0,...,0,2])}(q,y) + 7\cdot \chi_{(0;[2,0,...,0,2,0])}(q,y) \nonumber \\
& & + \ 9\cdot \chi_{(0;[3,0,...,0,2])}(q,y) + 9\cdot \chi_{(0;[2,0,...,0,3])}(q,y) \nonumber \\
& & + \  2\cdot\chi_{(0;[0,1,0,...,0,2,0])}(q,y) + 2\cdot\chi_{(0;[0,2,0,...,0,1,0])}(q,y)  \nonumber \\
& & + \ 2\cdot \chi_{(0;[0,1,0...,0,3])}(q,y)  +  2\cdot \chi_{(0;[3,0,...,0,1,0])}(q,y) \nonumber \\
& & + \  6\cdot \chi_{(0;[1,1,0,...,0,1,1])}(q,y) + \ {\cal O}(q^{7/2}) \ ,
\end{eqnarray}
which indeed matches exactly (\ref{E2}) to this order. 

\section{Embedding of $S_{N+1}$ in ${\rm U}(N)$}\label{app:embedding}

One can explicitly study the embedding of $S_{N+1}$ in ${\rm U}(N)$ (or for that matter, ${\rm O}(N)$).
 We consider $\mathbb{C}^{N+1}$ (or $\mathbb{R}^{N+1}$ in the case of ${\rm O}(N)$) with a set of holomorphic 
 orthonormal basis vectors $\vec{e}_i$ $(i=1,\ldots, (N+1))$. We consider the $N$-dimensional subspace 
 perpendicular to the vector $\vec{A}_0 =\frac{1}{\sqrt{N+1}}\, (\sum_{i=1}^{N+1}\vec{e}_i)$, which has 
 the following convenient orthonormal basis
\begin{eqnarray}
\vec{A}_1 &=& \frac{1}{\sqrt{N(N+1)}}\, \Bigl(\, \sum_{i=1}^{N}\vec{e}_i- N\vec{e}_{N+1}\Bigr) \nonumber \\
\vec{A}_2 &=& \frac{1}{\sqrt{N(N-1)}}\, \Bigl(\, \sum_{i=1}^{N-1}\vec{e}_i- (N-1)\vec{e}_{N}\Bigr) \nonumber \\
\vdots & & \vdots \\
\vec{A}_m &=& \frac{1}{\sqrt{(N-m+1)(N-m+2)}}\, \Bigl(\, \sum_{i=1}^{N-m+1}\vec{e}_i- (N-m+1)\vec{e}_{N-m+2}
\Bigr) \nonumber \\
\vdots & & \vdots \nonumber \\
\vec{A}_N &=& \frac{1}{\sqrt{2}}\, (\vec{e}_1-\vec{e}_2) \ . \nonumber
\end{eqnarray}
Then ${\rm U}(N)$ then acts by the usual complex rotations on this orthonormal basis $\{\vec{A}_m \}$.

We can realise the permutation group $S_{N+1}$ as a subgroup in the following way. $S_{N+1}$ has a 
natural permutation action on the basis $\vec{e}_i$ by permuting the indices. This action leaves 
$\vec{A}_0$ invariant, and thus the $N$-dimensional space orthogonal to it. 
In terms of the basis $\{\vec{A}_m \}$ we can write down the $S_{N+1}$ action explicitly. 
The permutation group is generated by the $N$ elementary transpositions of adjacent indices. 
Let us introduce the notation $T_m = (N-m+1 \, \, N-m+2)$ for $m=1,\ldots, N$ so that 
\be
T_m\cdot \vec{e}_{N-m+1}=\vec{e}_{N-m+2}\ ; \qquad T_m\cdot \vec{e}_{N-m+2}=\vec{e}_{N-m+1}\ ,
\ee
with $T_m$ leaving all other $\vec{e}_i$ unchanged. We then find that the nontrivial 
action of these generators on the  $\vec{A}_n$ is (for $m=1,\ldots, (N-1)$)
\be
T_m\cdot \vec{A}_{m}=-\alpha_m\,  \vec{A}_{m}+\beta_m\, \vec{A}_{m+1}\ ; \qquad 
T_m\cdot \vec{A}_{m+1} = \beta_m\, \vec{A}_{m}+\alpha_m \, \vec{A}_{m+1}\ ,
\ee
where
\be
\alpha_m=\frac{1}{(N-m+1)} \qquad \hbox{and} \qquad \beta_m=\frac{\sqrt{(N-m)(N-m+2)}}{(N-m+1)}\ ,
\ee
with $\alpha_m^2+\beta_m^2=1$, while all other basis vectors $\vec{A}_n$ are left unchanged.  
The remaining transposition, $T_N =(1 2)$, leaves all the $\vec{A}_n$ invariant except for 
$T_N\cdot\vec{A}_N = - \vec{A}_N$. Thus we see from this explicit construction that these transpositions 
(and thus the group they generate) are all unitary matrices (actually orthogonal matrices). Since 
$\det(T_m)=-1$ for all $m=1,\ldots, N$, they lie in ${\rm O}(N)$ (or ${\rm U}(N)$) rather than in
${\rm SO}(N)$ (or ${\rm SU}(N)$).  

\subsection{$S_{N+1}$ Singlet Multiplicities in ${\rm U}(N)$ Representations}\label{app:singlets}

The organisation of the vacuum character of the symmetric product in terms of characters of the 
coset theory relies on knowing the multiplicities of singlets of the symmetric group $S_{N+1}$ appearing in 
nontrivial representations of ${\rm U}(N)$. In principle, this is determined by the branching rules for the 
decomposition of a ${\rm U}(N)$ representation in terms of its $S_{N+1}$ subgroup. These branching rules are 
somewhat complicated and though, in principle, they can be extracted from the literature (see, e.g., 
\cite{King, Brown:2008ij, Brown:2010pb}) it is not very easy in practice. In this appendix we describe a 
rough and ready algorithm to determine the multiplicities  which is easy to use for representations with a small 
number of boxes and anti-boxes. This suffices for the counting of multiplicities to the order we check in this work. 
At higher orders one needs to augment the rules given below and the counting gets more involved. Presumably 
with a bit more work one can also write down generating 
functions for the general multiplicities along the lines mentioned for simple classes below. 

Consider $(N+1)$ variables $x_i$ with $i=1,\ldots, (N+1)$. As in the previous section, we can consider 
these to be coordinates in $\mathbb{C}^{N+1}$. We can then look at the $N$-dimensional subspace 
orthogonal to the hyperplane $\sum_i x_i=0$. Then, as described in the previous subsection, 
${\rm U}(N)$ has a natural action on this subspace with the independent combinations of the $x_i$ transforming 
in the fundamental representation. We also saw that the permutation group $S_{N+1}$ acts in the obvious way 
by permuting the indices $i$. The projection to $\sum_i x_i=0$ removes the singlet part. The remaining $N$ 
independent components transform in the standard or $N$ dimensional representation of $S_{N+1}$. 

Let us first consider the simplest case of completely symmetric representations of ${\rm U}(N)$, i.e., of the form 
$[\ell,0\ldots, 0]$ (or its complex conjugate $[0\ldots, 0, \ell]$). We want to determine the number of 
$S_{N+1}$ singlets in these representations. Thus we look at the $\ell$'th powers of $x_i$ subject to the 
condition $\sum_ix_i=0$. Since the permutation invariant combinations of any number of variables are
 generated by the power sums $s_m(x) = \sum_i x_i^m$, we simply have to count the number of different 
 ways that we can write terms of homogeneity $\ell$ from the products of $s_m(x)$, remembering that 
 $s_1(x)=0$. This number is given by the number of ways we can partition $\ell$ into sums of integers each 
 greater than one. Each such combination will be an inequivalent way of forming an $S_{N+1}$ singlet 
 from the symmetric powers of the fundamental.

If we denote the number of these singlets by $N(\ell)$, then we see that its generating function is given by
\be
\sum_{\ell=0}^{\infty}N(\ell)\, u^k=\prod_{n=2}^{\infty}\frac{1}{(1-u^n)}  \ .
\ee
Thus we have $N(2)=1$, $N(3)=1$, $N(4)=2$, etc. 
For the conjugate representation we clearly have the same answer. But since we will need to combine fundamentals 
and anti fundamentals we will denote the corresponding variable for the latter by $y_i$. While combining the two 
we need to keep in mind that we are not interested in the ${\rm U}(N)$ singlet representations formed when we 
tensor a box and an anti-box --- these are automatically $S_{N+1}$ singlets. Thus we will impose the condition that 
$\sum_i x_iy_i=0$. Therefore when we consider representations with the Dynkin labels $[\ell,0\ldots, 0, \bar\ell]$, 
we look  at all the symmetric combinations we can write down with $\ell$ $x_i$'s and $\bar\ell$ $y_i$'s, 
subject to the conditions $\sum_i x_i=\sum_i y_i =\sum_i x_iy_i=0$. 
Now the most general $S_{N+1}$ singlet combinations are $s_{m,n}(x,y) = \sum_i x_i^my_i^n$, 
subject to the conditions  $s_{1,0}(x,y) =s_{0,1}(x,y) =s_{1,1}(x,y)=0$. Thus we can again write down a generating 
function for the number of $S_{N+1}$-singlets $N(\ell,\bar\ell)$ in the representation $[\ell,0\ldots, 0,\bar\ell]$ as 
\be
\sum_{\ell,\bar\ell=0}^{\infty}\, N(\ell,\bar\ell)\, u_1^\ell \, \bar{u}_1^{\bar\ell} =\prod^{\prime}_{(n\geq0,m\geq0)}
\frac{1}{(1-u_1^n\, \bar{u}_1^{\, m})} \ ,
\ee
where the prime means that we exclude the pairs $(n,m)=(0,0),  (0,1), (1,0), (1,1)$.  

We can go on to consider more general representations involving antisymmetric tensors. Thus corresponding to the 
Dynkin label $[0,1,\ldots 0]$ we introduce a new variable $x_{[ij]}$. The square brackets denote antisymmetrisation in 
the enclosed indices; one should view this roughly as $x_i\wedge x_j$. However we note that since $i,j= 1,\ldots, (N+1)$ 
this does not by itself represent the antisymmetric tensor power of two fundamentals. In fact, the relation $\sum_ix_i=0$ 
implies that $\sum_ix_{[ij]}=\sum_j x_{[ij]} =0$. One can then verify that these $N$ constraints indeed reduces the 
number of independent components to that of a second rank antisymmetric tensor of ${\rm U}(N)$. 
When combining this with anti-fundamentals, since we project out the ${\rm U}(N)$ singlets we must also impose 
the constraints $\sum_i x_{[ij]} y_i = \sum_j x_{[ij]}y_j=0$. Once again we form $S_{N+1}$ invariant combinations by
 taking products of power sums subject to the constraints above as well as taking into account antisymmetry in the indices. 
There are now many possibilities. Thus, for  
instance, in the representation $[0,2,\ldots,0]$ we have one singlet combination $\sum_{i,j}x_{[ij]}^2$. And for 
$[3,1,\ldots,0]$ we have one combination $\sum_{i,j}x_{[ij]}x_i^2x_j$. For something more nontrivial like 
$[1,2,\ldots, 0,1]$ we have two singlets, namely, $\sum_{i,j}x_{[ij]}^2x_i^2y_i$ as well as $\sum_{i,j}x_{[ij]}^2x_i^2y_j$. 
In general, we can also have products of such invariants, such as $(\sum_{i,j}x_{[ij]}^2)(\sum_ky_k^2)$ as one of the three different $S_{N+1}$ singlet combinations in the representation $[0,2,0\ldots 0,2]$. 

The rules for including more general antisymmetric powers is similar. We introduce new variables 
$x_{[i_1,\ldots i_n]} \sim x_{i_1}\wedge\ldots\wedge x_{i_n}$,  and similarly $y_{[j_1, \ldots j_m]}$ for 
the anti boxes. We impose the constraints 
\be
\sum_{i_1}x_{[i_1,\ldots i_n]}=\sum_{j_1}y_{[j_1,\ldots j_m]}=0
\ee
as well as 
\be
\sum_k x_{[k, i_1,\ldots i_{n-1}]}\, y_{[k, j_1,\ldots j_{m-1}]}=0\ .
\ee
These arise, as before, from projecting out the extra degrees of freedom in these antisymmetric tensors and 
throwing away the ${\rm U}(N)$ singlet pieces. We write down all the elementary power sums of these variables 
which are non-vanishing after taking into account the constraints as well as the antisymmetry\footnote{There would be additional relations, when one considers representations with different types of antisymmetric tensors, to take into account the mixed symmetry. To the order to which we are checking the equality of characters, these additional relations do not play a role. For instance, in the table, for $B=6$, we need to use the mixed symmetry properties to rule out the potential singlet $\sum_{i,j,k}x_ix_{[jk]}x_{[ijk]}$ for the representation $[1,1,1,0\ldots,0]$. But this representation contributes in any case only at $O(q^{\frac{7}{2}})$.}. We then 
form products of these of the right homogeneity in the independent variables corresponding to the Dynkin labels of 
the ${\rm U}(N)$ representation. Thus a representation like $[4,8,0,2,\ldots 3,5]$ will have homogeneity 
$(4,8,2,3,5)$ respectively, in the variables $x_i, x_{[ij]}, x_{[ijkl]}, y_{[ij]}, y_i$, respectively. Using these rules one 
can compute the non-zero multiplicities $n(\Lambda)$ for the representations
with up to 6 boxes and anti-boxes (that contribute up to ${\cal O}(q^3)$). We find\footnote{We thank
Marco Baggio for helping us check these branching rules also explicitly.}
\begin{eqnarray}
B =2 : &\quad   & \bigl(\, {\tiny\yng(2)}\ ,0 \bigr) \ , \   \bigl(0, {\tiny\yng(2)}\, \bigr) \nonumber  \\
B =3 :&\quad  & \bigl(\, {\tiny\yng(3)}\ ,0 \bigr) \ , \   \bigl(0, {\tiny\yng(3)}\, \bigr) \ , \ 
\bigl(\, {\tiny\yng(2)}\ ,\tiny{\yng(1)}\, \bigr) \ , \  \bigl(\tiny{\yng(1)}\ , {\tiny\yng(2)}\, \bigr) 
 \nonumber \\
 B =4 :& \quad & 2\cdot \bigl(\, {\tiny\yng(4)}\ ,0 \bigr) \ ,   \ 2 \cdot \bigl(0, {\tiny\yng(4)}\, \bigr) \ , \ 
\bigl(\, {\tiny\yng(2,2)}\ ,0 \bigr) \ ,   \ \bigl(0, {\tiny\yng(2,2)}\, \bigr) \ ,  \nonumber \\
& & \ 
\bigl(\, {\tiny\yng(3)}\ ,\tiny{\yng(1)}\, \bigr) \ ,  \ \bigl(\tiny{\yng(1)}\ , {\tiny\yng(3)}\, \bigr) \ , \ 
2\cdot \bigl(\tiny{\yng(2)}\ , {\tiny\yng(2)}\, \bigr)  \nonumber 
\end{eqnarray}
\begin{eqnarray}
B=5: & &  2\cdot \bigl(\, {\tiny\yng(5)}\ ,0 \bigr) \ ,   \ 2 \cdot \bigl(0, {\tiny\yng(5)}\, \bigr) \ , \
\bigl(\, {\tiny\yng(4,1)}\ ,0 \bigr) \ ,   \ \bigl(0, {\tiny\yng(4,1)}\, \bigr) \ , \nonumber \\
& & \ \bigl(\, {\tiny\yng(3,2)}\ ,0 \bigr) \ ,   \ \bigl(0, {\tiny\yng(3,2)}\, \bigr) \ , \nonumber \\
& & \ 
2\cdot \bigl(\, {\tiny\yng(4)}\ ,\tiny{\yng(1)}\, \bigr) \ ,   \ 2 \cdot \bigl(\tiny{\yng(1)}\ , {\tiny\yng(4)}\, \bigr) \ , \
\bigl(\, {\tiny\yng(3,1)}\ ,\tiny{\yng(1)}\, \bigr) \ ,   \bigl(\tiny{\yng(1)}\ , {\tiny\yng(3,1)}\, \bigr) \ , \nonumber \\
& & \ 
\bigl(\, {\tiny\yng(2,2)}\ ,\tiny{\yng(1)}\,  \bigr) \ ,   \ \bigl(\tiny{\yng(1)}\, , {\tiny\yng(2,2)}\, \bigr) \ ,  \nonumber \\
& & \ 3 \cdot \bigl(\, {\tiny\yng(3)}\ ,\tiny{\yng(2)}\, \bigr) \ ,  \ 3\cdot  \bigl(\tiny{\yng(2)}\ , {\tiny\yng(3)}\, \bigr) \ , \ 
\bigl(\, {\tiny\yng(2,1)}\ ,\tiny{\yng(2)}\, \bigr) \ ,  \   \bigl(\tiny{\yng(2)}\ , {\tiny\yng(2,1)}\, \bigr) \label{5box}
\end{eqnarray}
\begin{eqnarray}
B=6: & \quad & 4\cdot \bigl(\, {\tiny\yng(6)}\ ,0 \bigr) \ ,   \ 4 \cdot \bigl(0, {\tiny\yng(6)}\, \bigr) \ , \ 
3\cdot \bigl(\, {\tiny\yng(4,2)}\ ,0 \bigr) \ ,   \ 3 \cdot \bigl(0, {\tiny\yng(4,2)}\, \bigr) \ , \nonumber \\
& &\  \bigl(\, {\tiny\yng(5,1)}\ ,0 \bigr) \ ,   \  \bigl(0, {\tiny\yng(5,1)}\, \bigr) \ , \ 
\bigl(\, {\tiny\yng(2,2,2)}\ ,0 \bigr) \ ,   \  \bigl(0, {\tiny\yng(2,2,2)}\, \bigr) \ , \nonumber \\
& &  \ 3\cdot \bigl(\, {\tiny\yng(5)}\ ,\tiny{\yng(1)}\, \bigr) \ ,   \ 3 \cdot \bigl(\tiny{\yng(1)}\ , {\tiny\yng(5)}\, \bigr) \ , \
2\cdot \bigl(\, {\tiny\yng(4,1)}\ ,\tiny{\yng(1)}\, \bigr) \ ,   \ 2 \cdot \bigl(\tiny{\yng(1)}\ , {\tiny\yng(4,1)}\, \bigr) \ , \nonumber \\
& & \ 2 \cdot  \bigl(\, {\tiny\yng(3,2)}\ ,\tiny{\yng(1)}\, \bigr) \ ,   \  2 \cdot  \bigl(\tiny{\yng(1)}\ , {\tiny\yng(3,2)}\, \bigr) \ ,  \nonumber \\
& & \ 6 \cdot  \bigl(\, {\tiny\yng(4)}\ ,\tiny{\yng(2)}\, \bigr) \ ,   \  6 \cdot  \bigl(\tiny{\yng(2)}\ , {\tiny\yng(4)}\, \bigr) \ , \
3 \cdot  \bigl(\, {\tiny\yng(2,2)}\ ,\tiny{\yng(2)}\, \bigr) \ ,   \  3 \cdot  \bigl(\tiny{\yng(2)}\ , {\tiny\yng(2,2)}\, \bigr) \ ,  
\nonumber \\
& & \  
2 \cdot  \bigl(\, {\tiny\yng(3,1)}\ ,\tiny{\yng(2)}\, \bigr) \ ,   \  2 \cdot  \bigl(\tiny{\yng(2)}\ , {\tiny\yng(3,1)}\, \bigr) \ , \
\bigl(\, {\tiny\yng(3,1)}\ ,\tiny{\yng(1,1)}\, \bigr) \ ,   \   \bigl(\,\tiny{\yng(1,1)}\ , {\tiny\yng(3,1)}\, \bigr) \ , \nonumber \\
& & \ 5\cdot  \bigl(\, {\tiny\yng(3)}\ ,\tiny{\yng(3)}\, \bigr) \ ,   \ 
2\cdot \bigl(\, {\tiny\yng(2,1)}\ ,\tiny{\yng(3)}\, \bigr) \ ,   \   2\cdot \bigl(\tiny{\yng(3)}\ , {\tiny\yng(2,1)}\, \bigr) \ ,
\bigl(\, {\tiny\yng(2,1)}\ ,\tiny{\yng(2,1)}\, \bigr) \ . \label{6box}
\end{eqnarray}
Here we have described the representations of ${\rm U}(N)$ in terms of pairs of Young diagrams (describing
the boxes and anti-boxes, respectively), and $B$ is the total number of boxes and anti-boxes.

\bibliographystyle{JHEP}

\end{document}